\documentclass{SciPost}

\usepackage{tcolorbox}
\usepackage{braket}
\usepackage{floatrow}
\usepackage{subcaption}
\usepackage{listings}
\usepackage{xcolor}
\usepackage{orcidlink}
\usepackage{url}
\usepackage{quantikz}

\newcommand{\code}[1]{\ilcode{\detokenize{#1}}}

\lstdefinelanguage{julia}{
  keywords={function, end, if, else, elseif, for, while, try, catch, struct, module, using, import, abstract, primitive, type, typealias, quote, let, begin, break, continue, return, macro, global, local, const},
  keywordstyle=\color{blue}\bfseries,
  sensitive=true,
  comment=[l]{\#},
  morestring=[b]",
  morestring=[b]',
  basicstyle=\ttfamily\footnotesize
}
\hypersetup{
    colorlinks,
    linkcolor={red!50!black},
    citecolor={blue!50!black},
    urlcolor={blue!80!black}
}

\usepackage{tikz}
\usetikzlibrary{arrows.meta,decorations.pathreplacing,positioning,calc,shadings}
\usepackage[bitstream-charter]{mathdesign}
\DeclareSymbolFont{usualmathcal}{OMS}{cmsy}{m}{n}
\DeclareSymbolFontAlphabet{\mathcal}{usualmathcal}

\fancypagestyle{SPstyle}{
\fancyhf{}
\lhead{\colorbox{scipostblue}{\bf \color{white} ~SciPost Physics Codebases }}
\rhead{{\bf \color{scipostdeepblue} ~Submission }}

\fancyfoot[C]{\textbf{\thepage}}
}

\definecolor{python-background}{HTML}{F6F8FA}
\definecolor{python-margin}{HTML}{91a3b5}
\definecolor{python-string}{HTML}{0A3069}
\definecolor{python-comment}{HTML}{59636E}
\definecolor{python-keywords}{HTML}{CF222E}

\lstdefinestyle{python}%
{ basicstyle=\ttfamily%
	, identifierstyle=%
	, commentstyle=\color{python-comment}%
	, stringstyle=\color{python-string}%
	, keywordstyle=\color{python-keywords}
	, columns=spaceflexible%
	, keepspaces=true%
	, showspaces=false%
	, showtabs=false%
	, showstringspaces=false%
}%

\lstdefinestyle{boxed}{
	style=python, %
	frame=trbL
	numberstyle=\small,
	frame=leftline,
	numbersep=7pt,
	framesep=5pt,
	framerule=2pt,
	xleftmargin=2pt,
	backgroundcolor=\color{python-background}, %
	rulecolor=\color{python-margin}, %
	breaklines=true, %
}

\definecolor{purple}{HTML}{98358c} 
\definecolor{backPurple}{HTML}{F7EBF4} 
\definecolor{cyan}{HTML}{0a7d91}
\definecolor{backCyan}{HTML}{D3E8ED}
\definecolor{backGrey}{HTML}{f3f4f5}
\definecolor{borderGrey}{HTML}{d1d5da}

\newcommand{\cbox}[1]{%
\tcbox[on line,%
       boxsep=1pt,%
       left=2pt,%
       right=2pt,%
       top=0pt,%
       bottom=0pt,%
       colback=backGrey,%
       colframe=borderGrey, %
       boxrule=0.5pt]{\color{purple}\texttt{#1}}}

\newcommand{\module}[1]{%
\tcbox[on line,%
       boxsep=1pt,%
       left=2pt,%
       right=2pt,%
       top=0pt,%
       bottom=0pt,%
       colback=backGrey,%
       colframe=borderGrey, %
       boxrule=0.5pt]{\color{cyan}\texttt{#1}}}

\newcommand{\ilcode}[1]{%
\tcbox[on line,%
       boxsep=1pt,%
       left=2pt,%
       right=2pt,%
       top=0pt,%
       bottom=0pt,%
       colback=backGrey,%
       colframe=borderGrey, %
       boxrule=0.5pt]{\texttt{#1}}}

\newcommand\atomtwin{%
\tcbox[on line,%
       boxsep=1pt,%
       left=2pt,%
       right=2pt,%
       top=0pt,%
       bottom=0pt,%
       colback=backGrey,%
       colframe=borderGrey, %
       boxrule=0.5pt]{\color{cyan}\texttt{AtomTwin}}}

\begin{document}

\pagestyle{SPstyle}

\begin{center}{\Large \textbf{\color{scipostdeepblue}{
				AtomTwin.jl: a physics-native digital twin framework for neutral-atom quantum processors
			}}}\end{center}

\begin{center}\textbf{
		Shannon Whitlock\,\orcidlink{0000-0002-6955-9326}\textsuperscript{1$\star$,2}
	}\end{center}

\begin{center}
	{\bf 1} European Center for Quantum Sciences, ISIS-CESQ (UMR7006), University of Strasbourg and CNRS\\
	{\bf 2} QPerfect SAS, 23 rue du Loess, Strasbourg, France
	\\[\baselineskip]
	$\star$ \href{mailto:whitlock@unistra.fr}{\small whitlock@unistra.fr}
\end{center}

\section*{\color{scipostdeepblue}{Abstract}}
\textbf{\boldmath{%
AtomTwin.jl is an open-source Julia package for developing and simulating quantum protocols, hardware configurations and building digital twins for neutral-atom quantum processors and related atomic quantum devices. AtomTwin operates between mathematical models and physical devices; modeling atoms, optical tweezers, laser fields, atomic motion, interactions, and noise processes natively from physical geometry and parameters, without requiring users to define Hamiltonians manually. The package provides hardware-level instruction sequences, high-performance solvers for coupled quantum and classical dynamics, and a ready-to-use model for ytterbium-171 atoms in an extensible framework designed to accommodate a greater variety of atomic species and hardware components in the future. This paper describes the software architecture, performance benchmarks against existing toolboxes, and a demonstrated end-to-end application: preparation of a logical Bell state in the $[[4,2,2]]$ error-detecting code with four $^{171}$Yb atoms in moveable tweezers.}}

\vspace{\baselineskip}

\noindent\textcolor{white!90!black}{%
	\fbox{\parbox{0.975\linewidth}{%
			\textcolor{white!40!black}{\begin{tabular}{lr}%
					\begin{minipage}{0.6\textwidth}%
						{\small Copyright attribution to authors. \newline
							This work is a submission to SciPost Physics Codebases. \newline
							License information to appear upon publication. \newline
							Publication information to appear upon publication.}
					\end{minipage} & \begin{minipage}{0.4\textwidth}
						                 {\small Received Date \newline Accepted Date \newline Published Date}%
					                 \end{minipage}
				\end{tabular}}
		}}
}


\vspace{10pt}
\noindent\rule{\textwidth}{1pt}
\tableofcontents
\noindent\rule{\textwidth}{1pt}
\vspace{10pt}


\section{Introduction}

\label{sec:intro}

Designing and validating neutral-atom quantum processors requires navigating a large space of interdependent choices, including atomic species, qubit/level assignment, trap geometry, laser configurations, pulse sequences, and error minimization strategies. In other engineering disciplines, such complexity is typically managed through hardware-realistic digital simulation. For example, modern classical processors are designed using electronic design automation (EDA) tools that integrate device models, circuit behavior, and system-level verification. An analogous simulation capability — one that connects physical device parameters to instruction-level quantum behavior — is required for quantum processors, but the underlying physics (coupled internal states, atomic motion, driven dynamics, stochastic noise, etc.) fundamentally differs from classical systems, motivating integrated, physics-native design and simulation frameworks that connect physical processes and system dynamics directly to instruction-level behavior.

General-purpose open-source quantum simulation libraries such as QuTiP~\cite{johansson2012,lambert2024} and QuantumOptics.jl~\cite{kramer2018} provide flexible frameworks for simulating open and closed quantum systems. However, they are formulated at the level of user-defined Hamiltonians and dissipators, with no native representation of physical hardware. As a result, modeling a realistic device requires manually translating atomic structure, laser couplings, and noise processes into mathematical operators, a process that is tedious, error-prone, and slow to prototype at scale.

On the other side, tools such as Pulser~\cite{silverio2022} and Bloqade~\cite{bloqade2023} are designed for programming neutral-atom experiments with a focus on pulse-level control of effective many-body Hamiltonians. These frameworks provide convenient abstractions for driving Rydberg systems, but rely on simplified models (typically two-level atoms with parametrized interactions) and offer limited support for extending the underlying physics to include realistic level structures, spatially varying fields required for site-resolved and gate-based control, or atomic motion, an important error source for current and near-term neutral-atom processors. Complementary packages address specific sub-problems: ARC~\cite{robertson2021} and PairInteraction~\cite{weber2017} compute atomic properties and interaction potentials, while AtomECS~\cite{chen2021}, PyLCP~\cite{eckel2022}, and atomSmltr~\cite{weill2026} simulate laser cooling and atomic motion, but do not address the full coherent many-body dynamics of quantum processors.

\atomtwin{} is designed to help bridge this gap. It provides a physics-native simulation environment for neutral atoms in which hardware components are specified directly, and system dynamics are constructed automatically from these descriptions. By combining realistic atomic structure, spatially resolved fields, time-dependent control and accurate noise processes within a unified framework, \atomtwin{} aims to enable end-to-end simulation of neutral-atom quantum systems at a component level, from atomic physics to multiqubit instructions. Its goal is integrate known physics into a consistent hardware-native simulation environment, enabling reproducibility and more direct comparisons across studies through a shared representation of device-level physics. It is intended for the design, development, and validation of hardware-specific quantum protocols, used alongside or prior to experimental implementation. The same framework could serve as the basis for a quantum digital twin: a continuously updated virtual model of a specific processor, maintained in synchrony with hardware through calibration and control data, enabling closed-loop comparison between simulated and measured behavior. The current implementation covers the core physics: semiclassical dynamics, realistic level structures, spatially resolved fields, Rydberg interactions, noise modeling, and quantum process tomography, as well as dynamical qubit architectures including programmable trap geometries and in-sequence atom transport. It is designed to be extensible: more atomic species, interaction types, noise processes, and detector types can be contributed as additional system objects allowing the platform to evolve with the needs of the community.

This paper is structured as follows: Section~\ref{sec:package-description} describes the package design and user-facing API. Section~\ref{sec:implementation} details the numerical implementation: the directed acyclic graph (DAG) based compilation pipeline and highly optimized solvers for mixed quantum and classical dynamics. Section~\ref{sec:benchmark} validates accuracy and compares performance against other commonly used toolboxes on two benchmarks: two-level Rabi oscillations with dephasing, and collective Rydberg blockade dynamics. Section~\ref{sec:application-example} showcases an end-to-end simulation of logical Bell state preparation protocol in the $[[4,2,2]]$ error-detecting code on four ytterbium-171 atoms.

\newpage
\section{Package description}
\label{sec:package-description}

\subsection{Overview and design philosophy}

\atomtwin{} is a Julia package for physics-level modeling of neutral-atom quantum processors and related atomic quantum devices. Its central design principle is that simulations are built from physical components: atomic species with realistic internal level structures, optical tweezers, and laser beams with specified characteristics, rather than from abstract Hamiltonians, while still giving the user control over which physical processes to include. The system dynamics are constructed automatically from user specified physical processes, reducing the gap between the physical intuition behind a device and its computational model. This approach reflects the intended use case: a hardware or control system engineer should be able to describe the relevant physical components in terms that map directly onto standard experimental parameters, without first translating that description into a Hamiltonian by hand. It also means that modifying a component, e.g., changing a beam power, adding a noise source, or even switching to a different atomic species, propagates consistently through the simulation without requiring the user to reconstruct the underlying operators.

A typical \atomtwin{} workflow proceeds in three main stages, illustrated in Figure~\ref{fig:workflow}. First, a \cbox{System} is defined by specifying the physical components of the device, including atoms with internal level structure and motional degrees of freedom, optical fields and trapping geometries, interaction terms, dissipative processes, and detector models. In parallel, a \cbox{Sequence} defines a time-ordered control program consisting of instructions such as pulses, delays, and tweezer movements.

Simulations are initiated by calling \cbox{play}, which takes a system and sequence as input. This triggers a compilation step that constructs a \cbox{SimulationJob}: a compiled, time-resolved representation of the system dynamics with pre-allocated memory, and concrete solver objects. This functional structure mirrors the way a hardware control system typically operates: \atomtwin{} is designed to be used alongside, or as a precursor to the real hardware control stack, so that the sequence abstraction mirrors the structure of hardware control programs, facilitating eventual translation into hardware instructions. For multi-shot simulations, the cost of compilation is partially amortized: \cbox{play} reuses the existing job and performs incremental updates via \cbox{recompile!}, modifying only shot-dependent elements such as noise realizations or parameter variations. The job is then executed by the \module{AtomTwin.Dynamiq} backend, which advances the classical and quantum degrees of freedom in time using automatically selected solvers (e.g. Schrödinger, master equation, wavefunction Monte-Carlo and semiclassical variants), while recording detector outputs.

\usetikzlibrary{arrows.meta, positioning, calc}

\definecolor{hdrBlue}{RGB}{31,97,141}
\definecolor{bkgBlue}{RGB}{214,234,248}
\definecolor{hdrSlate}{RGB}{52,73,94}
\definecolor{bkgSlate}{RGB}{235,239,243}
\definecolor{hdrTeal}{RGB}{0,105,92}
\definecolor{bkgTeal}{RGB}{212,239,235}
\definecolor{hdrGray}{RGB}{75,88,98}
\definecolor{bkgGray}{RGB}{242,244,246}

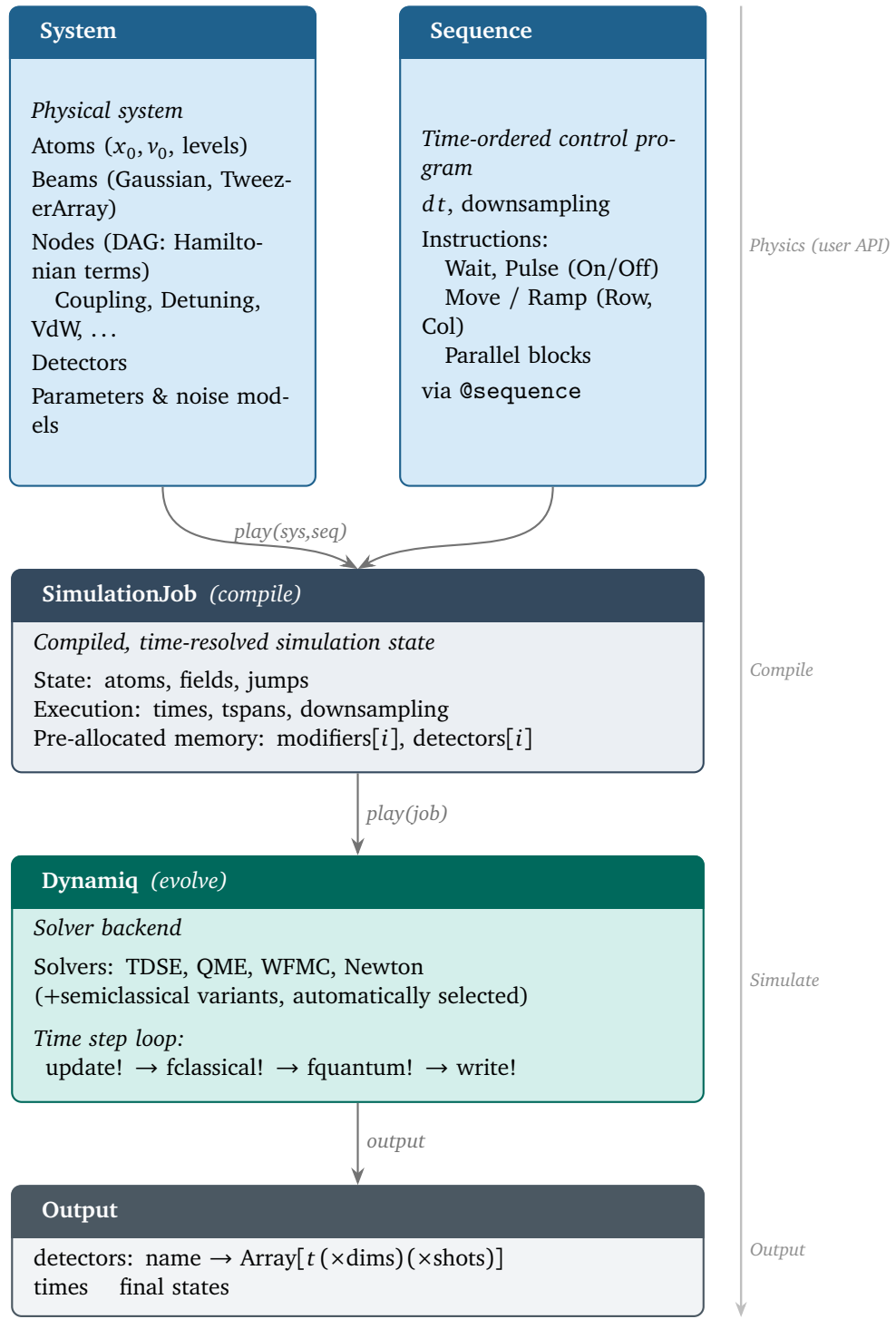
\begin{figure*}[htbp]
    \centering
    \hspace*{2cm}%
    \begin{tikzpicture}[
        box/.style    = {draw, thick, rounded corners=4pt,
                         align=left, font=\small, inner sep=9pt},
        sysbox/.style = {box, fill=bkgBlue,  draw=hdrBlue,  minimum height=7.0cm},
        seqbox/.style = {box, fill=bkgBlue,  draw=hdrBlue,  minimum height=7.0cm},
        jobbox/.style = {box, fill=bkgSlate, draw=hdrSlate},
        dynbox/.style = {box, fill=bkgTeal,  draw=hdrTeal},
        outbox/.style = {box, fill=bkgGray,  draw=hdrGray},
        arr/.style    = {-{Stealth[scale=1.1]}, thick, color=black!55},
        lbl/.style    = {font=\footnotesize\itshape, color=black!55},
      ]
    
      \node[sysbox, text width=3.8cm] (sys) {%
        \rule{0pt}{12pt}\\
        \textit{Physical system}\\[3pt]
        Atoms ($x_0, v_0$, levels)\\[2pt]
        Beams (Gaussian, TweezerArray)\\[2pt]
        Nodes (DAG: Hamiltonian terms)\\
        \quad Coupling, Detuning, VdW, $\dots$\\[2pt]
        Detectors\\[2pt]
        Parameters \& noise models%
      };
      \begin{scope}
        \clip[rounded corners=4pt] (sys.south west) rectangle (sys.north east);
        \fill[hdrBlue] ([yshift=-22pt]sys.north west) rectangle (sys.north east);
      \end{scope}
      \node[text=white, font=\small\bfseries, anchor=mid west]
        at ([yshift=-11pt, xshift=9pt]sys.north west) {System};
    
      \node[seqbox, text width=3.8cm, right=1.2cm of sys] (seq) {%
        \rule{0pt}{12pt}\\
        \textit{Time-ordered control program}\\[3pt]
        $dt$, downsampling\\[3pt]
        Instructions:\\
        \quad Wait, Pulse (On/Off)\\
        \quad Move / Ramp (Row, Col)\\
        \quad Parallel blocks\\[3pt]
        via \texttt{@sequence}%
      };
      \begin{scope}
        \clip[rounded corners=4pt] (seq.south west) rectangle (seq.north east);
        \fill[hdrBlue] ([yshift=-22pt]seq.north west) rectangle (seq.north east);
      \end{scope}
      \node[text=white, font=\small\bfseries, anchor=mid west]
        at ([yshift=-11pt, xshift=9pt]seq.north west) {Sequence};
    
      \node[jobbox, text width=9.4cm,
            below=1.2cm of $(sys.south)!0.5!(seq.south)$] (job) {%
        \rule{0pt}{12pt}\\
        \textit{Compiled, time-resolved simulation state}\\[4pt]
        State: atoms, fields, jumps\\
        Execution: times, tspans, downsampling\\
        Pre-allocated memory: modifiers[$i$], detectors[$i$]%
      };
      \begin{scope}
        \clip[rounded corners=4pt] (job.south west) rectangle (job.north east);
        \fill[hdrSlate] ([yshift=-22pt]job.north west) rectangle (job.north east);
      \end{scope}
      \node[text=white, font=\small\bfseries, anchor=mid west]
        at ([yshift=-11pt, xshift=9pt]job.north west)
        {SimulationJob\enspace\normalfont\small\itshape(compile)};
    
      \node[dynbox, text width=9.4cm, below=1.2cm of job] (dyn) {%
        \rule{0pt}{12pt}\\
        \textit{Solver backend}\\[4pt]
        Solvers: TDSE, QME, WFMC, Newton\\
        (+semiclassical variants, automatically selected)\\[6pt]
        \textit{Time step loop:}\\
        \enspace update! $\to$ fclassical! $\to$ fquantum! $\to$ write!%
      };
      \begin{scope}
        \clip[rounded corners=4pt] (dyn.south west) rectangle (dyn.north east);
        \fill[hdrTeal] ([yshift=-22pt]dyn.north west) rectangle (dyn.north east);
      \end{scope}
      \node[text=white, font=\small\bfseries, anchor=mid west]
        at ([yshift=-11pt, xshift=9pt]dyn.north west)
        {Dynamiq\enspace\normalfont\small\itshape(evolve)};
    
      \node[outbox, text width=9.4cm, below=1.2cm of dyn] (out) {%
        \rule{0pt}{12pt}\\
        detectors:\enspace name $\to$ Array$[t\,(\times\mathrm{dims})\,(\times\mathrm{shots})]$\\
        times \quad final states%
      };
      \begin{scope}
        \clip[rounded corners=4pt] (out.south west) rectangle (out.north east);
        \fill[hdrGray] ([yshift=-22pt]out.north west) rectangle (out.north east);
      \end{scope}
      \node[text=white, font=\small\bfseries, anchor=mid west]
        at ([yshift=-11pt, xshift=9pt]out.north west) {Output};
    
      \draw[arr] (sys.south) to[out=-90, in=150]
        node[left, lbl, pos=1.0, yshift=15pt] {play(sys,seq)} (job.north);
      \draw[arr] (seq.south) to[out=-90, in=30] (job.north);
      \draw[arr] (job.south) -- node[right, lbl] {play(job)} (dyn.north);
      \draw[arr] (dyn.south) -- node[right, lbl] {output} (out.north);
    
      \coordinate (rx) at ([xshift=0.5cm]seq.east);
      \draw[-{Stealth[scale=0.9]}, color=black!25, line width=0.9pt]
        (rx |- sys.north) -- (rx |- out.south);
      \node[font=\scriptsize\itshape, color=black!45, anchor=west]
        at (rx |- sys.center) {Physics (user API)};
      \node[font=\scriptsize\itshape, color=black!45, anchor=west]
        at (rx |- job.center) {Compile};
      \node[font=\scriptsize\itshape, color=black!45, anchor=west]
        at (rx |- dyn.center) {Simulate};
      \node[font=\scriptsize\itshape, color=black!45, anchor=west]
        at (rx |- out.center) {Output};    
    \end{tikzpicture}
\caption{\atomtwin{} workflow. A \cbox{System} (atoms, beams, nodes, detectors) and a \cbox{Sequence} (time-ordered instructions) are compiled into a \cbox{SimulationJob} by \ilcode{play} and then dispatched to the \module{Dynamiq} solver, which advances the classical and quantum degrees of freedom step by step and writes detector outputs.}
\label{fig:workflow}
\end{figure*}

\subsection{System model}

\subsubsection{Atoms and internal structure}

Atoms in \atomtwin{} are defined by their internal level structure. Levels can be generic (\cbox{Level}), fine-structure sublevels (\cbox{FineLevel}, parametrized by $J$, $m_J$, and the Landé $g$-factor), or hyperfine sublevels (\cbox{HyperfineLevel}, parametrized by $F$, $m_F$ and $g_F$). Groups of sublevels sharing a common $J$ or $F$ can be organized into \cbox{FineManifold} or \cbox{HyperfineManifold} objects, from which selection rules and Clebsch--Gordan coefficients are computed automatically when adding couplings. Any subset of levels can be selected for a given simulation, enabling efficient reduction of the Hilbert space to the most physically relevant degrees of freedom.

Built-in species types (\cbox{Ytterbium171Atom}, \cbox{Rubidium87Atom}, \cbox{Strontium88Atom}, \cbox{Potassium39Atom}) carry default mass and nuclear spin; polarizability models and transition data are provided for \cbox{Ytterbium171Atom}, based on experimental data~\cite{hohn2023}; the other built-in species support level structure and mechanics as well as manually specified wavelength-dependent polarizabilities. A generic \cbox{Atom} type supports species-agnostic simulations. Each atom carries a classical center-of-mass position and velocity, enabling semiclassical simulations in which the internal quantum state and the motional state of each atom are propagated simultaneously.

\subsubsection{Beams and tweezer arrays}

Optical fields are represented as beam objects. A \cbox{GaussianBeam} models an axis-aligned focused Gaussian beam characterized by wavelength, waist radius, and power; a \cbox{GeneralGaussianBeam} extends this to elliptical beams with arbitrary propagation direction, position dependent phase and polarization. A \cbox{PlanarBeam} represents a plane wave with spatially uniform intensity, a fixed wavevector, and polarization. Two-dimensional acousto-optic deflector (AOD) tweezer arrays are supported via the \cbox{TweezerArray} type, which parameterizes a rectangular grid of Gaussian beams by AOD drive frequencies and amplitudes in two orthogonal directions.

Polarization is specified as a Cartesian vector $\boldsymbol{\epsilon} = (\epsilon_x,\epsilon_y,\epsilon_z)$ in the laboratory frame and converted to the spherical basis ($q=-1,0,1$) for computing angular-dependent couplings. Polarization impurity is handled through a \cbox{MixedPolarization} type, which models a dominant polarization component with a parametric admixture of an orthogonal contamination; for \cbox{MixedPolarization} the contamination amplitude is sampled stochastically at compile time and the resulting vector is normalized.

\subsubsection{Physical processes}

Physical processes are added to a system via a set of functions that construct the corresponding Hamiltonian or Lindblad terms. \cbox{add\_coupling!} adds a coherent laser coupling between two levels with a specified Rabi frequency; when given manifold arguments, selection rules and Clebsch--Gordan coefficients are applied automatically to distribute coupling strength across the relevant sublevel pairs. It also accepts an optional beam parameter which controls the spatially dependent complex amplitude of the coupling(s). \cbox{add\_detuning!} adds a single-level energy shift, and \cbox{add\_zeeman\_detunings!} applies field-dependent Zeeman shifts to all sublevels of a manifold. Dissipative processes are added via \cbox{add\_decay!} (spontaneous emission between levels, with relative decay rates derived from Clebsch--Gordan coefficients when manifolds are specified) and \cbox{add\_dephasing!} (pure dephasing on selected levels). Two-body interactions between pairs of levels are added via \cbox{add\_interaction!} (constant interaction) and \cbox{add\_vdwinteraction!} ($C_6/r^6$ distance dependent interaction).

Mechanical forces on the atoms are handled automatically when a beam is added to the System: the beam geometry and the scalar dynamic polarizability at the beam wavelength determine the position-dependent trapping potential and any state-dependent light-shifts. This allows the trapping and coupling fields to be specified and modified independently.

\subsection{Simulation methods}

\subsubsection{Quantum dynamics}

\module{AtomTwin.Dynamiq} supports three built-in quantum dynamics simulation methods. The method is selected automatically based on whether or not dissipative processes are present in the system and the \ilcode{density\_matrix} keyword of \cbox{play}.

\begin{itemize}
    \item \textit{Schrödinger equation} (default for unitary dynamics, \ilcode{density\_matrix=false}): deterministic evolution of a state vector $|\psi(t)\rangle$ under the system Hamiltonian. This is the fastest method, appropriate for closed systems or short-time dynamics.
    \item \textit{Lindblad master equation} (\ilcode{density\_matrix=true}): evolution of a density matrix $\rho(t)$ according to
    \begin{equation}
        \dot{\rho} = -\frac{i}{\hbar}[H, \rho] + \sum_k \left( L_k \rho L_k^\dagger - \frac{1}{2}\{L_k^\dagger L_k, \rho\} \right),
        \label{eq:lindblad}
    \end{equation}
    where $L_k$ are Lindblad jump operators encoding spontaneous decay and dephasing. This method is in principle exact for Markovian open systems but scales as $d^2$ with Hilbert space dimension $d$, which becomes expensive for multi-atom systems.
    \item \textit{Monte Carlo wavefunction (MCWF)} method~\cite{dalibard1992} (\ilcode{density\_matrix=false} when dissipative processes are present): individual quantum trajectories are propagated under a non-Hermitian effective Hamiltonian with stochastic quantum jumps drawn from the jump operators. Averaging over many trajectories recovers the density matrix result, but individual trajectories can also give physical insight into single-shot measurement outcomes. The method scales as $d$ per trajectory and is therefore preferred for larger systems, especially when stochastic averaging over noise realizations is required. Multi-shot execution is parallelized automatically across available threads.
\end{itemize}

\subsubsection{Semiclassical motion}

When any atom carries a non-zero velocity or the atom carries polarizability data at the wavelength of a beam in the system, \atomtwin{} automatically engages semiclassical mode and evolves the classical center-of-mass motion concurrently with the quantum internal state. Initial positions and velocities can be fixed or drawn from a distribution (e.g., \cbox{MaxwellBoltzmann} for thermal atoms, \cbox{GaussianPosition} for positional disorder). The equations of motion are integrated using Newton’s law, with forces derived from the optical potential of the tweezers and laser beams. This semiclassical treatment enables realistic modeling of Doppler shifts, position-dependent Rabi frequencies, finite Rydberg blockade strength, and motional dephasing, all of which are significant error sources in current neutral-atom processors. The semiclassical approximation is valid when the thermal de Broglie wavelength is small compared to the spatial scale of the optical potential (typically satisfied for atoms in optical tweezers at temperatures of order $1\,\mu\text{K}$ and above). Genuine quantum motional effects such as motional quantization, sideband structure, recoil suppression, and motional entanglement are outside the present scope, but could be treated by adding explicit quantum motional states to the model. Classical-only simulations (Newton's method) are also possible when no \ilcode{initial\_state} is set on the \cbox{System} or passed to \cbox{play}, which is useful for applications such as tweezer sorting and array rearrangement.

\subsubsection{Sequence and instruction set}

Control of the quantum system is expressed through a \cbox{Sequence}: a time-ordered list of instructions executed by \cbox{play}. Available instructions include \cbox{Pulse} (apply a coupling for a fixed duration, optionally with a shaped complex amplitude profile), \cbox{Wait} (idle period), \cbox{On}/\cbox{Off} (toggle couplings), and tweezer control instructions (\cbox{MoveCol}, \cbox{MoveRow}, \cbox{RampCol}, \cbox{AmplCol}, \cbox{FreqCol}, etc.) for AOD-driven atom transport and handover protocols. Sequences are assembled using the \cbox{@sequence} macro, which supports native Julia control flow (loops, conditionals) within the sequence body, enabling programmatic construction of complex instruction patterns. This design accommodates complex experimental protocols, e.g. dynamical decoupling, rearrangement sequences, parametric sweeps, that are most naturally expressed as programs rather than as flat lists of instructions.

\subsection{Parameters, noise, and error budgeting}

Physical quantities in \atomtwin{} can be represented as numbers or symbolic \cbox{Parameter} objects. Parameters carry a name, a default value, and an optional standard deviation for shot-to-shot disorder. Parameters compose arithmetically into \cbox{ParametricExpression} trees, which are resolved at compile time for each shot. This enables systematic sweeps over physical variables (e.g., Rabi frequency, detuning, temperature) and stochastic sampling of calibration errors without rebuilding the whole system model. The separation between model structure and numerical values allows hardware-realistic noise studies: users can sample errors or perform parameter scans without manually constructing a new system each time.

Correlated laser phase noise is modeled via \cbox{LaserPhaseNoiseModel}, which parameterizes the one-sided frequency-noise power spectral density (PSD) as the sum of a Gaussian servo bump and a power-law background,
\begin{equation}
    S_\nu(f) = A_G \exp\!\left(-\frac{(f - f_0)^2}{2\sigma^2}\right) + A_\mathrm{pl}\, f^{\,\alpha},
    \label{eq:psd}
\end{equation}
where $f_0$ and $\sigma$ are the center frequency and width of the servo bump, and $\alpha$ is the power-law exponent (e.g.\ $\alpha = 0$ for white noise).
Each shot samples a unique time-domain phase noise trajectory from this spectrum and attaches it to a coupling. The phase noise trajectory is continuous across the entire pulse sequence, so correlations in the noise are preserved between gates and idle periods, allowing users to quantify the effects of non-Markovian error sources directly. 

\subsection{Measurement and analysis}

\subsubsection{Final state}

Passing \cbox{final\_state=true} to the simulator returns the full quantum state at the end of the simulation. For closed-system simulations this is the state vector; for open-system (Lindblad) simulations it is the density matrix; for trajectory-based simulations it is the ensemble of per-trajectory state vectors. This provides direct access to the final state for downstream analysis, including state tomography, entanglement measures, or custom post-processing beyond the built-in detectors.

\subsubsection{Detectors}

Observables are registered as detector specifications before running a simulation. \cbox{PopulationDetectorSpec} records the time-resolved population of a specified level; \cbox{CoherenceDetectorSpec} records an off-diagonal density matrix element; \cbox{MotionDetectorSpec} records atomic positions and velocities; \cbox{FieldDetectorSpec} records the complex amplitude envelope of a coupling or beam. Multiple detectors can be registered on a single system and are evaluated at the end of each time step with minimal overhead.

\subsubsection{Quantum process tomography}

The \cbox{process\_tomography} function performs single-qubit quantum process tomography by simulating the system's action on a complete set of input states and reconstructing the Choi matrix, Pauli transfer matrix (PTM), and a Kraus operator representation of the resulting quantum channel. This enables direct computation of process fidelities, identification of dominant error mechanisms, and comparison with ideal gate targets.

\section{Implementation}
\label{sec:implementation}

\subsection{Package structure}

\atomtwin{} is implemented as two cooperating layers within a single Julia package. The user-facing layer provides atomic physics models, physical component constructors, and orchestration logic described in Section~\ref{sec:package-description}. The numerical engine, \module{AtomTwin.Dynamiq}, implements the time integrators, Hamiltonian, jump operators, and detectors. The two layers communicate through a typed interface: \atomtwin{} constructs a directed acyclic graph (DAG) of physics nodes encoding a particular system and compiles it into a \cbox{SimulationJob} with preallocated operators and state arrays. \cbox{Dynamiq} then integrates the time evolution using highly optimized solvers. Julia's multiple dispatch allows \atomtwin{} to overload \module{Dynamiq} constructors so that \cbox{Parameter} and \cbox{ParametricExpression} objects are accepted wherever ordinary numbers appear, with the parametric logic resolved before the solver loop begins rather than inside it.

Keeping the two layers separate has practical advantages. Changes to the physical model (adding a noise source, modifying an atomic species, changing beam geometry) are expressed entirely in the \atomtwin{} layer and propagate automatically to the compiled operators; the \module{Dynamiq} solver code is unmodified. Conversely, performance work in \module{Dynamiq} (integrators, parallelism strategy, operator storage format) does not touch the user API.

\subsection{System representation and DAG compilation}

\subsubsection{Node graph}

A \cbox{System} is represented internally as an ordered sequence of \cbox{AbstractNode} objects that implicitly encodes a directed acyclic graph (DAG): dependencies between nodes are resolved topologically, while nodes with no mutual dependencies are processed in insertion order. Each node encodes one physical component or process: \cbox{BeamNode} for an optical field, \cbox{CouplingNode} or \cbox{NoisyCouplingNode} for a coherent laser coupling (with or without laser phase noise), \cbox{PlanarCouplingNode} for a plane-wave coupling with position-dependent phase, \cbox{DetuningNode} for an energy shift, \cbox{DecayNode} for spontaneous emission, and \cbox{InteractionNode} for a two-body Rydberg interaction.

Every node participates in a two-method resolution. \code{_resolve_node_default(x)} is called at build time and extracts default numerical values from any \cbox{Parameter} objects the node contains. \code{_resolve_node_value(x, param_values, rng)} is called at compile and recompile time with a concrete parameter sample and a random-number generator, enabling stochastic draws and per-shot parametric variation. Quantities that vary across shots, e.g., Rabi frequencies drawn from calibration uncertainty, shot-to-shot phase noise, polarization contamination amplitudes, are represented as \cbox{Parameter} or \cbox{ParametricExpression} objects and automatically resolved through this protocol.

A minimal system definition illustrates the structure:

\begin{lstlisting}[language=julia]
using AtomTwin

# define qubit levels from the 3P0 nuclear-spin manifold
g = HyperfineLevel(1//2, 0, -1//2; label = "1S0, m_F=-1/2")
e = HyperfineLevel(1//2, 0, +1//2; label = "3P0, m_F=1/2")

# atom with those levels; tweezer passed at System construction
atom    = Ytterbium171Atom(; levels = [g, e])
tweezer = GaussianBeam(λ = 759e-9, w0 = 0.8e-6, P = 0.5e-3)
system  = System(atom, tweezer)

# add a resonant clock drive and spontaneous decay
coupling = add_coupling!(system, atom, g => e, 2π * 2e3)
decay = add_decay!(system, atom, e => g, 2π * 0.0076)

# inspect the node list
for node in system.nodes
    println(typeof(node))
end
\end{lstlisting}

\noindent The \ilcode{system.nodes} list reflects insertion order. The compilation pipeline resolves execution order via topological sort of the DAG, so the user-facing insertion order does not need to respect dependencies.

\atomtwin{} also supports native Hilbert space reduction through the \ilcode{maxoccupations} keyword of \cbox{System}. A constraint of the form \ilcode{(level, n)} restricts the basis to states in which the specified level is occupied by at most $n$ atoms across the array. The most common case is \ilcode{(r, 1)}, which limits the Rydberg population to at most one atom at a time: in the strong blockade regime this is enforced by physics, and the constraint simply removes unreachable basis states from the computation, reducing the Hilbert space dimension and the cost of every operator--vector product. For a five-atom array of three-level atoms ($\{|0\rangle, |1\rangle, |r\rangle\}$) with at most one Rydberg excitation, the unrestricted dimension $3^5 = 243$ is reduced to $2^5 + 5 \times 2^4 = 112$.

\subsubsection{Compilation pipeline}

Calling \ilcode{compile(system, seq)}, where \ilcode{seq} is a \cbox{Sequence} (Section~\ref{sec:package-description}), converts the node graph into a \cbox{SimulationJob}. Compilation proceeds in three ordered phases. First, all \cbox{BeamNode} objects are resolved: any parametric beam properties (wavelength, waist, power) are fixed to concrete values for this shot, producing concrete \cbox{AbstractBeam} objects. Position-dependent field amplitudes are not evaluated at this stage; they are recomputed at each solver timestep from the atoms' current positions. Second, atoms are initialized with positions, velocities, and scalar polarizabilities: if a polarizability model is defined for the atomic species, polarizabilities are computed automatically at the wavelength of each beam object in the system; otherwise the user must supply them explicitly. Third, all remaining nodes are compiled in topologically sorted order, constructing the numerical Hamiltonian and Lindblad operators at the atoms' initial positions.

The output is a \cbox{SimulationJob}: a struct holding the time-dependent Hamiltonian and jump operators as sparse \cbox{Op} arrays, preallocated state vectors, detector buffers, and parameter metadata. For multi-shot execution, \code{recompile!(job, sys; rng, kwargs...)} updates these numerical arrays in place for each new shot, sampling parameters and noise realizations without allocating new memory. The separation between structure (compile once) and values (recompile per shot) makes large parameter sweeps and multishot runs as efficient as possible.

A parametric Rabi frequency illustrates how parameters propagate through to \ilcode{play}:

\begin{lstlisting}[language=julia]
# declare a Rabi frequency with shot-to-shot Gaussian disorder
Ω = Parameter(:Ω, 2π * 5e3; std=2π * 0.1e3)

coupling = add_coupling!(system, atom, g => e, Ω)

# each shot draws Ω from a normal distribution with mean 5 kHz, sigma 0.1 kHz
out = play(system, seq; initial_state = g, shots = 200)

# fix Ω at a specific sweep point
out2 = play(system, seq; initial_state = g, Ω = 2π * 6e3)
\end{lstlisting}

\noindent Each shot in the multi-shot call draws a fresh sample from a normal distribution and recompiles only the affected operators in place; no user intervention is required. Passing a keyword argument matching the parameter name overrides sampling with a fixed value, enabling deterministic sweeps over physical variables.

Laser phase noise is handled by \cbox{NoisyCouplingNode}, which synthesizes a per-shot time-domain phase trajectory from a \cbox{LaserPhaseNoiseModel} (see Section~\ref{sec:package-description}). The phase trajectory is computed via spectral shaping of white noise and integrated in the frequency domain; each shot receives an independent realization of the stochastic phase $\phi(t)$ that modulates the coupling amplitude as $\Omega \to \Omega\, e^{i\phi(t)}$ throughout the sequence.

\subsection{Quantum propagator}

\subsubsection{Operator representation}

\module{Dynamiq} stores Hamiltonians and jump operators in a sparse \cbox{Op} type, optimized for quantum dynamics solvers. An \cbox{Op} holds the nonzero matrix elements as a flat list of (row, column, value) triplets grouped into forward and reverse (conjugate) parts; this representation avoids redundant storage of Hermitian structure and maps efficiently onto the memory access pattern of the propagator. At compile time, \cbox{Op} objects are constructed once and remain static during evolution; time-dependence is modeled through time-dependent complex coefficients stored in the physics objects, which pre-multiply the \cbox{Op} at each step (with conjugates applied to reverse parts). This separation of static operators and dynamic coefficients allows rapid updates without recompiling operators. Separate \cbox{Op} objects are created for the time-independent part (static light shifts, detunings, interaction energies).

\subsubsection{Time evolution}

State evolution over a single time step $[t, t+\delta t]$ is handled by \ilcode{fquantum!} using an operator-splitting strategy. The unitary part is propagated by a Taylor expansion of the time-evolution operator (in units $\hbar=1$),
\begin{equation}
    e^{-iH\delta t} \approx \sum_{n=0}^{N} \frac{(-iH\delta t)^n}{n!},
    \label{eq:taylor}
\end{equation}
where the order $N$ is configurable (default: 4). This expansion is applied directly to the state vector via repeated sparse matrix--vector products; no matrix exponential is formed. For the master equation solver, Lindblad dissipator terms are applied as a first-order Euler step after the unitary part. This splitting is accurate when the dissipation rates $\Gamma_k$ are small compared to the coherent energy scales, which is the case in neutral-atom qubit experiments where gate times are typically much shorter than the coherence time $T_2$ by construction. An $N$-th order expansion has a local truncation error of $\mathcal{O}((\Omega\delta t)^{N+1})$; the first-order Euler step for the dissipator introduces an additional error of $\mathcal{O}((\Gamma\delta t)^2)$, and the operator-splitting (Lie--Trotter) error is $\mathcal{O}(\Gamma\Omega\delta t^2)$. For the step sizes used in practice ($\Omega\,\delta t \sim 10^{-2}$) the truncation error is typically small compared to the gate errors and observables of interest, as confirmed in Section~\ref{sec:benchmark}.

For the MCWF method, the non-Hermitian effective Hamiltonian $H_\mathrm{eff} = H - \frac{i\hbar}{2}\sum_k L_k^\dagger L_k$ replaces $H$ in Eq.~\eqref{eq:taylor}. After each step the norm $\langle\psi|\psi\rangle$ is compared to a threshold drawn uniformly at the start of each trajectory; when the norm falls below the threshold, a quantum jump is applied by selecting a jump operator $L_k$ with probability proportional to $\langle\psi|L_k^\dagger L_k|\psi\rangle$ and renormalizing the state~\cite{dalibard1992}.

\subsection{Semiclassical motion}

When atoms carry non-zero velocities or atomic polarizabilities are defined for the beams in the system, the quantum and classical degrees of freedom are co-integrated at each time step. The classical update \ilcode{fclassical!} advances positions and velocities using an Euler step with forces derived from the optical potential of all active tweezer and beam fields. In the classical \ilcode{newton} solver, the per-atom update loop is parallelized above a configurable atom-count threshold. For all solvers, multi-shot execution parallelizes trajectories over available threads via \ilcode{Threads.@threads}, and this shot-level parallelization takes precedence over atom-state parallelization.

\subsection{Sequence construction}

A \cbox{Sequence} is a time-ordered list of \cbox{AbstractInstruction} objects paired with a fixed time step \ilcode{dt} and an optional downsampling factor for detector outputs. Instructions such as \cbox{Pulse}, \cbox{Wait}, \cbox{On}, \cbox{Off}, and AOD transport commands (\cbox{MoveCol}, \cbox{RampCol}, etc.) are appended with \ilcode{push!(seq, instruction)}. Multiple instructions can be started simultaneously by wrapping them in a \cbox{Parallel} instruction, which dispatches all of its arguments at the same time step. This interface provides direct, imperative control over sequence construction. This is the appropriate interface when instructions are generated algorithmically, for example, when a compiled control routine produces a variable-length list of pulses at runtime:

\begin{lstlisting}[language=julia]
seq = Sequence(dt; downsample=10)

for (coupling, duration) in zip(couplings, durations)
    push!(seq, Pulse(coupling, duration))
    push!(seq, Wait(gap))
end
\end{lstlisting}

For sequences that are fixed at definition time, the \ilcode{@sequence} macro provides a more readable syntax. The macro rewrites each top-level function call inside a \ilcode{begin...end} block to a \ilcode{push!} call on a pre-declared \cbox{Sequence}, while preserving native Julia control flow (loops, conditionals, and local variables) so that complex instruction patterns can be expressed without leaving Julia:

\begin{lstlisting}[language=julia]
N_pi = 8
seq = Sequence(dt)

@sequence seq begin
    # initial π/2 pulse
    Pulse(coupling, π/2 / Ω)
    # dynamical decoupling: N_pi equally spaced π pulses
    for k in 1:N_pi
        Wait(tau)
        Pulse(coupling, π / Ω)
    end
    Wait(tau)
    # final π/2 pulse
    Pulse(coupling, π/2 / Ω)
end
\end{lstlisting}

\noindent Both interfaces build the same \cbox{Sequence} data structure and are fully interchangeable.

\subsection{Execution and parallelism}

Calling \ilcode{play(sys, seq; shots=N)} is the entry point for running a simulation. Internally, \ilcode{play} calls \ilcode{compile} to build the \cbox{SimulationJob}, then dispatches to the selected solver (state vector, master equation, or MCWF) for each shot. For multi-shot runs above a configurable threshold (default: 4 shots), execution is parallelized across available threads using \ilcode{Threads.@threads}; each thread operates on an independent copy of the state and detector arrays to avoid contention. A memory check before launching threads prevents over-subscription on machines with limited per-core memory; if the estimated footprint exceeds available RAM, execution falls back to sequential mode with a warning.

For a parametric sweep, e.g. varying detuning or pulse area across a grid, performance-sensitive code can call \ilcode{compile} once and then use the two-argument \ilcode{play(job, sys)} to reuse the compiled structure, calling \ilcode{recompile!} between sweep points to update only the affected operators in place. Parameter names are passed as keyword arguments directly:

\begin{lstlisting}[language=julia]
Δ = Parameter(:Δ, 0.0)
add_detuning!(system, atom, e, Δ)

job = compile(system, seq; initial_state = g)

results = map(Δ_range) do Δ_val
    recompile!(job, system; Δ = Δ_val)
    play(job, system; shots = 200)
end
\end{lstlisting}

\noindent The simpler \ilcode{play(system, seq; $\Delta$=$\Delta$\_val)} call compiles from scratch at each sweep point, which is adequate when the sweep is small or compile time is negligible relative to simulation time.

\noindent Detectors are evaluated at the end of each time step with negligible additional overhead; the \cbox{SimulationJob} carries preallocated detector buffers that are filled in place during integration and returned as named arrays at the end of \ilcode{play}. As a consequence, the output time grid starts at $\delta t$ rather than $t = 0$, i.e., detectors record the state after each step. This convention differs from some other simulation engines that return $t = 0$ as the first output point.

Time integration uses a fixed step size by default, though per-instruction step sizes and downsampling factors can be specified for flexibility. The fixed-step enables compile-time pre-allocation of detector arrays and simplifies synchronization between the two integrators in mixed quantum-classical simulations, but requires extra care to ensure numerical stability and convergence. Typically atomic motion is slow compared to the internal quantum dynamics, so convergence of the quantum part is typically sufficient to ensure overall accuracy. For now, convergence must be verified by the user, typically by varying $\delta t$ until the observable of interest no longer changes.

\section{Benchmarks}
\label{sec:benchmark}

To evaluate the performance of \atomtwin{} we benchmark against two widely used simulation packages: \module{QuantumOptics.jl} and \module{QuTiP}. These two tools represent the underlying numerical engines of the broader ecosystem: \module{Pulser}~\cite{silverio2022} uses \module{QuTiP} as its quantum simulation backend (at the time of writing), and \module{Bloqade}~\cite{bloqade2023} delegates time evolution to \module{DifferentialEquations.jl}, also used by \module{QuantumOptics.jl}. We also consider two specific test cases: coherent and dissipative single-qubit dynamics (Hilbert space dimension $d = 2$), and collective oscillations of $N$ interacting atoms in the Rydberg blockade regime ($d = 2^N$, $N = 2$--$8$). The first provides an accuracy reference against the closed-form optical Bloch equations; the second tests $N$-scaling under a stiff many-body Hamiltonian.

Timings measure raw simulation execution and exclude system construction and job compilation, reflecting the workload where many similar simulations are run in sequence, such as stochastic trajectories, tomography protocols, or parameter sweeps. All timings are minimum wall-clock times over repeated runs on an AMD Ryzen 7 PRO 7840U (8 physical cores, 16 threads, 5.1\,GHz boost, 32\,GB RAM) running Julia~1.11 (\atomtwin{} and \module{QuantumOptics.jl}) and Python~3 with \module{QuTiP}~5.2.3. \atomtwin{} uses a fixed time step $\delta t$ set per benchmark; \module{QuantumOptics.jl} and \module{QuTiP} use adaptive integrators (Dormand--Prince 5 and VODE/Adams, respectively) with default tolerances and output on a user-supplied time grid, at the same temporal resolution as \atomtwin{}. Because AtomTwin uses a fixed-step integrator while the reference packages use adaptive solvers, exact performance comparisons depend on the choice of timestep and tolerances and results are only indicative. For MCWF, trajectories are parallelized in all benchmarks: \atomtwin{} natively uses \ilcode{Threads.@threads}; \module{QuantumOptics.jl} has no native trajectory parallelism, so we wrap the trajectory loop with \ilcode{Threads.@threads} (shared-memory, same as \atomtwin{}); \module{QuTiP} distributes trajectories via \ilcode{multiprocessing} (separate processes) using the \ilcode{"map":"parallel"} option.

\subsection{Benchmark 1: Two-level Rabi oscillations with dephasing}
\label{sec:benchmark-rabi}

\subsubsection{Physical setup and motivation}

A two-level atom with a ground state $|g\rangle$ and an excited state $|e\rangle$, driven by a resonant field of amplitude $\Omega$, undergoes periodic oscillation of the population between the two levels — the Rabi oscillation. This is a fundamental single-qubit operation and a standard calibration measurement across all atom-based platforms. Including spontaneous decay $|e\rangle \to |g\rangle$ at rate $\Gamma \ll \Omega$ damps the oscillation amplitude slowly, and the full open-system dynamics admit an exact closed-form solution via the optical Bloch equations, providing an analytical accuracy reference for all three integration methods benchmarked here.

Here we benchmark the dynamics of a two-level atom ($|g\rangle$, $|e\rangle$) driven at $\Omega/2\pi = 1\,\text{MHz}$ with decay $\Gamma/2\pi = 0.5\,\text{kHz}$ ($\Gamma/\Omega = 5 \times 10^{-4}$). The simulation covers $T = 1\,\text{ms}$ (1000 Rabi periods); \atomtwin{} uses $\delta t = 10\,\text{ns}$ (100 steps per Rabi period). Three methods are benchmarked per engine: the Schrödinger equation ($\Gamma = 0$, reference $P_e(t) = \sin^2(\Omega t / 2)$), the Lindblad master equation, and MCWF (100 trajectories). Accuracy is the maximum absolute deviation of $P_e(t)$ from the exact solution: the undamped Rabi formula for the Schrödinger equation, and the full optical Bloch equation solution for the master equation and MCWF, over the final Rabi period.

\begin{figure*}[ht]
    \centering
    \includegraphics[width=\textwidth]{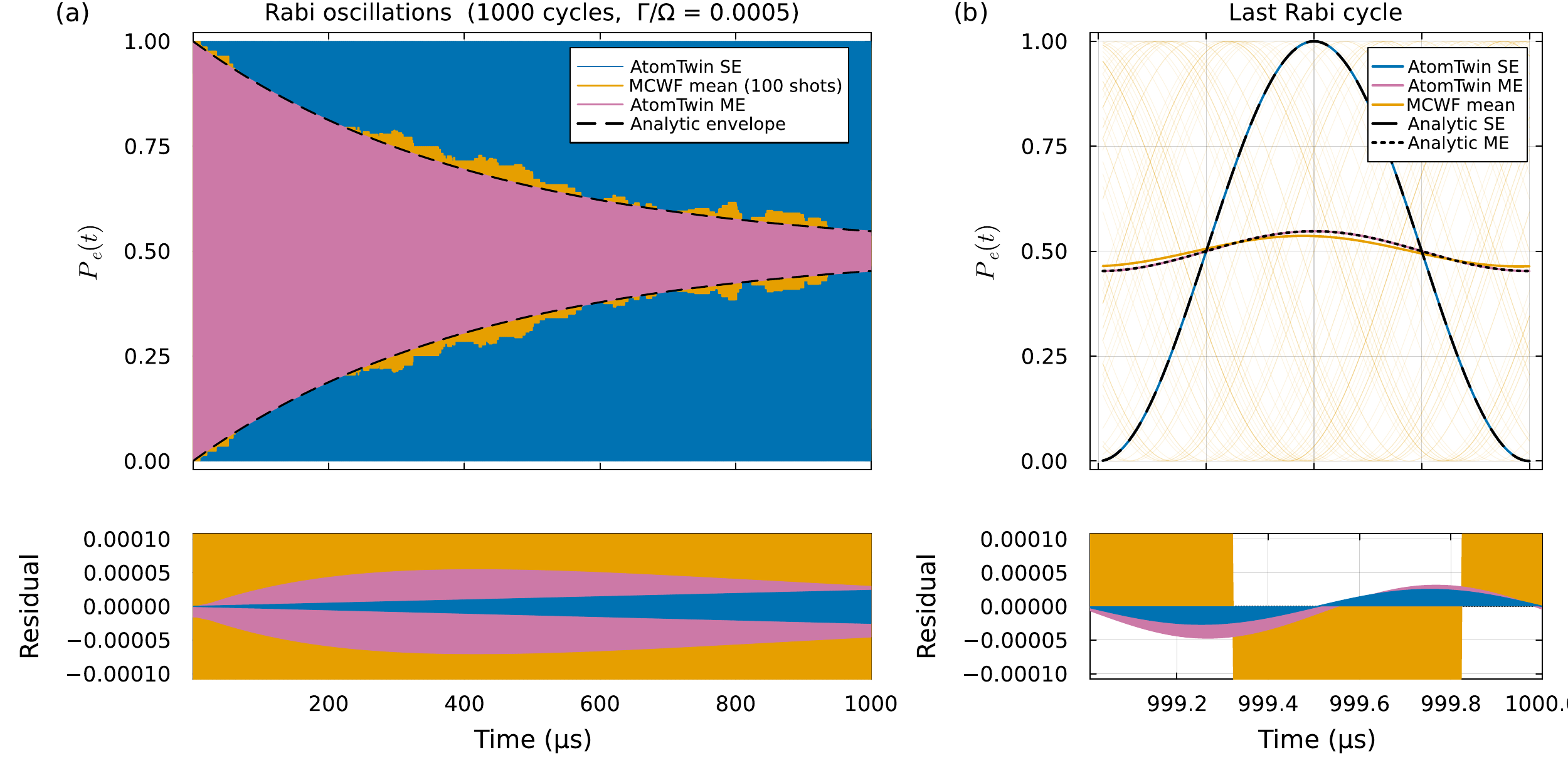}
    \caption{Rabi oscillation benchmark ($\Omega/2\pi = 1\,\text{MHz}$, $\Gamma/2\pi = 0.5\,\text{kHz}$, 1000 Rabi cycles, $\delta t = 10\,\text{ns}$, \atomtwin{}).
    \textbf{(a)} Excited-state population $P_e(t)$ over the full simulation: Schrödinger equation (dark blue), master equation (medium pink), and MCWF mean over 100 trajectories (light orange); Dashed black lines are the analytical envelope derived from the optical Bloch equations. Rabi oscillations are not resolved on the scale of the graph. \textbf{(b)} Zoom on the last Rabi cycle, with the analytical Schrödinger (dashed) and master-equation (dotted) solutions overlaid; all three methods track the analytic references to within visual resolution. Individual MCWF trajectories shown at reduced opacity.
    \textbf{Bottom panels} Residuals (simulated minus analytic).}
    \label{fig:benchmark-rabi}
\end{figure*}

\subsubsection{Results}

\begin{table}[ht]
\centering
\caption{Rabi oscillation benchmark: maximum deviation from the analytical solution over the last Rabi period, and wall-clock time (minimum over 10 runs). \atomtwin{} uses a fixed step $\delta t = 10\,\text{ns}$; \module{QuantumOptics.jl} uses adaptive Dormand--Prince integration; \module{QuTiP} uses adaptive VODE/Adams with default tolerances. MCWF uses 100 trajectories. Threaded rows use 16 threads: \atomtwin{} and \module{QuantumOptics.jl} (user-implemented) both use \ilcode{Threads.@threads} (shared memory); \module{QuTiP} uses \ilcode{multiprocessing} on 16 CPUs (separate processes). Errors for the MCWF methods are dominated by shot noise rather than numerical errors and so they are omitted from the table.}
\label{tab:benchmark-rabi}
\begin{tabular}{l l r r}
\hline
\textbf{Engine} & \textbf{Method} & \textbf{max\,$|$err$|$} & \textbf{Time (ms)} \\
\hline
\atomtwin{}         & Schrödinger (unitary)        & $2.6 \times 10^{-5}$ &      4.1 \\
                    & master equation               & $4.7 \times 10^{-5}$ &      8.8 \\
                    & MCWF (100 traj., serial)      & --- &    606.7 \\
                    & MCWF (100 traj., 16 threads)  & ---                  &    123.4 \\
\hline
\module{QuantumOptics.jl} & Schrödinger (unitary)        & $5.8 \times 10^{-4}$ &     16.1 \\
                           & master equation               & $8.8 \times 10^{-5}$ &     32.4  \\
                           & MCWF (100 traj., serial)      & --- &   3254.9   \\
                           & MCWF (100 traj., 16 threads)  & ---                  &   2451.9  \\
\hline
\module{QuTiP} (Python)   & Schrödinger (unitary)           & $2.0 \times 10^{-3}$ &     511.7  \\
                           & master equation                  & $3.4 \times 10^{-4}$ &     695.3 \\
                           & MCWF (100 traj., sequential)     & --- &  44\,450.7  \\
                           & MCWF (100 traj., 16 threads)        & ---                  &   8\,714.8  \\
\hline
\end{tabular}
\end{table}

All engines produce results consistent with the analytical references (Table~\ref{tab:benchmark-rabi}).

\noindent\textbf{Schrödinger equation.}
\atomtwin{} achieves $\max|$err$| = 2.6 \times 10^{-5}$ at $\delta t = 10\,\text{ns}$, a factor of $\sim 20$ below \module{QuantumOptics.jl} ($5.8 \times 10^{-4}$), while running $4\times$ faster ($4.1\,\text{ms}$ vs $16.1\,\text{ms}$). The accuracy advantage is consistent with \atomtwin{}'s dedicated Taylor propagator for Hamiltonian evolution and memory optimizations. \atomtwin{}'s fixed-step propagator pre-allocates all state arrays at compile time and accumulates detector outputs as scalars in place. \module{QuTiP} is more than two orders of magnitude slower and achieves $\max|$err$| = 2.0 \times 10^{-3}$ at default tolerances.

\noindent\textbf{Master equation.}
\atomtwin{} achieves $\max|$err$| = 4.7 \times 10^{-5}$ and runs in $8.8\,\text{ms}$, compared to $8.8 \times 10^{-5}$ and $32.4\,\text{ms}$ for \module{QuantumOptics.jl} — comparable accuracy at $3.7\times$ lower cost. This is notable given that \atomtwin{} uses only a first-order Euler step for the dissipator; at $\delta t = 10\,\text{ns}$ the dissipative contribution is not the limiting source of error ($\Gamma/\Omega = 5 \times 10^{-4}$). \module{QuTiP} achieves $\max|$err$| = 3.4 \times 10^{-4}$ and is more than two orders of magnitude slower, likely due to Python interpreter overheads.

\noindent\textbf{MCWF.}
\module{QuantumOptics.jl} returns full state trajectories from each \ilcode{mcwf} call, and the associated allocation cost grows with thread count rather than shrinking: at 16 threads the wall-clock time falls only from $3254\,\text{ms}$ to $2452\,\text{ms}$ ($1.3\times$ speedup), suggesting that memory management overhead limits good parallel scaling for this problem. \atomtwin{} writes detector scalars to pre-allocated arrays at each timestep; sequential execution completes in $607\,\text{ms}$ ($5.4\times$ faster than \module{QuantumOptics.jl}), and with 16 threads the time drops to $123\,\text{ms}$ ($20\times$ relative to \module{QuantumOptics.jl} on the same number of threads). The MCWF max\,$|$err$|$ of $\sim 3 \times 10^{-2}$ is dominated by shot noise from 100 trajectories in a weakly dissipative regime (few quantum jumps per trajectory). \module{QuTiP}'s \ilcode{mcsolve} distributes trajectories via \ilcode{multiprocessing}: sequential execution takes $44.5\,\text{s}$ and 16 CPUs reduce this to $8.7\,\text{s}$ ($5.1\times$ speedup), but this is still ${\sim}14\times$ slower than \atomtwin{} on a single thread or $\sim 70\times$ slower on multiple threads.

\subsection{Benchmark 2: Collective Rydberg Rabi oscillations in the blockade regime}
\label{sec:benchmark-blockade}

\subsubsection{Physical setup and motivation}

When $N$ two-level atoms are driven by a common resonant field and interact via a strong pairwise interaction $V \gg \Omega$, doubly-excited configurations are energetically suppressed (Rydberg blockade). As a result, the dynamics predominantly occupy the zero- and single-excitation manifold, within which the system behaves as an effective two-level system formed by the collective ground state and the symmetric single-excitation ($W$) state. The corresponding Rabi frequency is enhanced to $\sqrt{N}\,\Omega$, characteristic of the Rydberg superatom regime~\cite{dudin2012}. Collective coherent oscillations of this type were among the first experimental signatures of blockade physics~\cite{urban2009,gaetan2009} and underpin Rydberg-mediated entangling gates in neutral-atom platforms.

For the second benchmark we consider the system described by a quantum master equation with the Hamiltonian
\begin{equation}
H = \frac{\hbar\Omega}{2} \sum_{i} \sigma^x_i + \hbar V \sum_{i < j} n_i n_j,
\label{eq:blockade-H}
\end{equation}
where $\sigma^x_i$ is the Pauli-$X$ operator on site $i$ and $n_i = |r_i\rangle\langle r_i|$ is the Rydberg-state projector, and Lindblad jump operators $J_i = \sqrt{\gamma}\, n_i$, corresponding to pure dephasing of the Rydberg state and causing transitions out of the symmetric subspace. We use parameters $\Omega/2\pi = 1\,\mathrm{MHz}$, $V/2\pi = 100\,\mathrm{MHz}$, $\gamma/2\pi = 250\,\mathrm{kHz}$, and total evolution time $T = 1\,\mu\mathrm{s}$, with all atoms initialized in $|g\rangle$. In practice the pairwise interaction strength is distance-dependent, $V_{ij} = C_6/r_{ij}^6$; here we adopt a uniform value $V$ to avoid divergences and to better isolate the $N$-scaling of the solvers. Simulating this system in the full Hilbert space requires dimension $2^N$, and the interaction term introduces large energy scales that lead to numerical stiffness.

The interaction term produces eigenvalues up to $V N(N-1)/2$, but the dynamics remain predominantly confined to the zero- and single-excitation manifold. Empirically, we find that accurate integration requires a step size $\delta t = 1/(25\sqrt{N}\,V/2\pi)$. All engines output on a common time grid ($\sim 100$ points over $T$).

Additionally, \ilcode{\atomtwin{}} can enforce the blockade constraint at compile time by restricting evolution to the zero- and single-excitation manifold via \ilcode{maxoccupations=[(r,\,1)]}. This removes doubly-excited states entirely, eliminating the associated stiffness and yielding substantial speedups. For $V/\Omega = 100$, the resulting approximation incurs an error comparable to the ME integration errors; SE errors are substantially smaller (${\sim}10^{-8}$, see Table~\ref{tab:benchmark-blockade-scaling}).

\begin{figure*}[ht]
    \includegraphics[width=\textwidth]{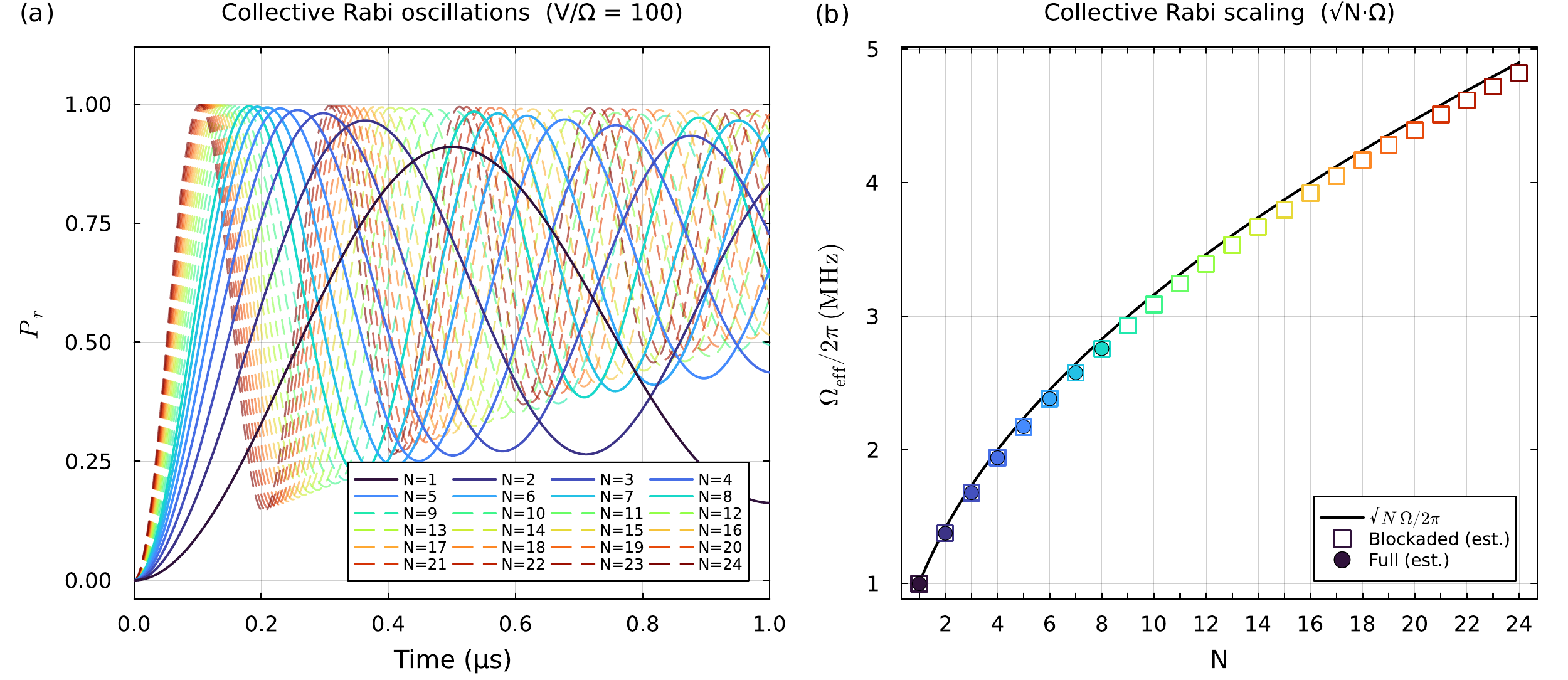}
    \caption{Collective Rydberg Rabi oscillations in the blockade regime ($\Omega/2\pi = 1\,\text{MHz}$, $V/\Omega = 100$, $\gamma/2\pi = 250\,\text{kHz}$), simulated using the full master equation including dephasing. \textbf{(a)} Total Rydberg population $P_r(t) = \langle\sum_i n_i\rangle$ over one single-atom Rabi period for $N = 1$--$8$ atoms in the full Hilbert space ($d = 2^N$, solid lines) and $N = 1$--$24$ in the blockaded subspace ($d = N+1$, dashed lines), color-coded by $N$. The blockaded subspace reproduces the full-space dynamics to high accuracy. \textbf{(b)} Effective Rabi frequency $\Omega_\mathrm{eff}/2\pi$ vs $N$ extracted from the first maximum of $P_r(t)$, for both the full space (circles) and blockaded subspace (squares), compared to the undamped $\sqrt{N}\,\Omega$ prediction (solid line). The frequencies estimated from the simulations fall slightly below the undamped prediction, which we attribute to the effect of dephasing.}
    \label{fig:benchmark-blockade}
\end{figure*}

\subsubsection{Results}

Results are summarized in Tables~\ref{tab:benchmark-blockade-scaling}--\ref{tab:benchmark-blockade-subspace}.

\begin{table*}[ht]
\centering
\caption{Rydberg blockade benchmark: wall-clock time (milliseconds) vs $N$, full Hilbert space ($d = 2^N$). MCWF uses 100 trajectories; \atomtwin{} and \module{QuantumOptics.jl} parallelize MCWF with \ilcode{Threads.@threads} on 16 threads; \module{QuTiP} uses \ilcode{"map":"parallel"} on 16 CPUs. SE accuracy $\max|P_r^\text{AT}-P_r^\text{QO}| \approx 10^{-8}$; ME accuracy $\approx 2\times10^{-4}$ (see text). \module{QuTiP} MCWF at $N=8$ was not completed due to the long runtime.}
\label{tab:benchmark-blockade-scaling}
\begin{tabular}{r | r r r | r r r | r r r}
\hline
 & \multicolumn{3}{c|}{\atomtwin{}} & \multicolumn{3}{c|}{\module{QuantumOptics.jl}} & \multicolumn{3}{c}{\module{QuTiP}} \\
$N$ & SE & ME & MCWF & SE & ME & MCWF & SE & ME & MCWF \\
\hline
2 & 0.4  & 1.4   & 19    & 1.0  & 1.8   & 50    & 20   & 27     & 7550   \\
3 & 1.5  & 9.3   & 49    & 2.5  & 16    & 211   & 32   & 51     & 10322  \\
4 & 4.0  & 61    & 463   & 12   & 118   & 763   & 57   & 90     & 13838  \\
5 & 7.4  & 291   & 1003  & 16   & 680   & 1259  & 109  & 1086   & 19864  \\
6 & 21   & 1829  & 2590  & 38   & 5206  & 3331  & 228  & 1932   & 31128  \\
7 & 56   & 18626 & 6799  & 86   & 54664 & 8221  & 452  & 56136  & 234562 \\
8 & 133  & 454114 & 17195 & 183  & 761195 & 21974 & 1035 & 373120 & ---    \\
\hline
\end{tabular}
\end{table*}

\noindent\textbf{Accuracy.}
\module{QuantumOptics.jl} with adaptive step control (Dormand--Prince 5, default tolerances) is used as the reference, which we confirmed converges to \atomtwin{} results as $\delta t \to 0$. For the fixed step size used for benchmarking, deviations $\max|P_r^\text{AT} - P_r^\text{QO}|$ are around $10^{-8}$ for the SE across $N = 2$--$8$. For the \atomtwin{} ME, the dominant source of error is the first-order Euler step of Lindblad dissipators which is more important in this benchmark than the previous one ($\gamma/\Omega = 0.25$ compared to $\Gamma/\Omega = 5\times 10^{-4}$). Accuracy of the SE and MCWF methods are unaffected. The blockaded-subspace results agree with the full-space reference to within $(\approx 1 \times 10^{-4}$, comparable to the numerical integration errors.

\noindent\textbf{Schrödinger equation performance.}
\atomtwin{} is up to $3\times$ faster for the SE than \module{QuantumOptics.jl} across $N = 2$--$8$ (Table~\ref{tab:benchmark-blockade-scaling}). \module{QuTiP} is ${\sim}14\times$ slower at $N = 4$ similar to what was observed in Benchmark~1. 

\noindent\textbf{Master equation performance.}
\atomtwin{} is consistently $2-3\times$ faster for the ME than \module{QuantumOptics.jl}, with the advantage growing with $N$, possibly attributable to lower allocation cost in \atomtwin{}'s fixed-step propagator, and pre-allocation of state arrays at compile time. \module{QuTiP} is ${\sim}1.5\times$ slower than \atomtwin{} at $N = 4$, but faster than \module{QuantumOptics.jl} at this system size. This contrasts with the large performance gap observed in Benchmark~1; the relative performance of \module{QuTiP} improves with Hilbert space dimension, consistent with fixed per-call overheads becoming negligible relative to the growing per-step numerical work.

\noindent\textbf{MCWF.}
Both Julia engines parallelize MCWF trajectories with \ilcode{Threads.@threads} (16 threads); \atomtwin{} is $1.6\times$ faster than \module{QuantumOptics.jl} at $N=4$ ($463\,\text{ms}$ vs $763\,\text{ms}$). \module{QuTiP} uses \ilcode{"map":"parallel"} on 16 CPUs and is ${\sim}30\times$ slower than \atomtwin{} at $N=4$ ($13.8\,\text{s}$ vs $463\,\text{ms}$).

\noindent\textbf{Blockaded subspace (\atomtwin{} only).}
Restricting evolution to the $N+1$-dimensional singly-excited subspace gives dramatic speedups, shown for the ME solver in Table~\ref{tab:benchmark-blockade-subspace}. At $N = 8$ the ME speedup is ${\sim}24\,000\times$ relative to the full space. Two effects contribute: the density matrix scales as $d^2$, so the dimension reduction from $2^N$ to $N+1$ alone accounts for a factor of $(2^N/(N+1))^2$; additionally, the blockaded Hamiltonian has $O(N)$ nonzero elements compared to $O(N \cdot 2^N)$ in the full space, so each operator-vector product is substantially cheaper. The SE speedup is more modest because the state vector scales as $d$ rather than $d^2$. A further speedup is possible by increasing the time step: in the blockaded subspace the stiff $V n_i n_j$ terms are absent, so $\delta t$ can be set by the slower Rabi scale $1/(\sqrt{N}\,\Omega)$ rather than $1/(\sqrt{N}\,V)$, potentially yielding an additional factor of $V/\Omega$ in step count. The blockaded subspace enables ME and MCWF simulations up to $N = 24$ in seconds or less, whereas the full-space ME becomes intractable above $N \approx 8$. Many gate protocols involve at most one or two simultaneous Rydberg excitations, so \ilcode{maxoccupations=[(r,\,1)]} or \ilcode{[(r,\,2)]} captures the relevant dynamics with negligible approximation error.

\begin{table}[ht]
\centering
\caption{\atomtwin{} wall-clock time (milliseconds) in the blockaded subspace ($d = N+1$, \ilcode{maxoccupations=[(r,\,1)]}), for even $N$. MCWF uses 100 trajectories on 16 threads. The ME speedup relative to the full Hilbert space ($d = 2^N$) is shown where the full-space benchmark is available; dashes indicate that the full-space ME is not feasible in the available time.}
\label{tab:benchmark-blockade-subspace}
\begin{tabular}{r r | r r r | r}
\hline
$N$ & $d$ & SE (ms) & ME (ms) & MCWF (ms) & ME speedup \\
\hline
 2 &  3 &  0.3 &   0.7 &    13 & $2\times$ \\
 4 &  5 &  1.1 &   3.2 &    36 & $19\times$ \\
 6 &  7 &  2.8 &   8.0 &    63 & $229\times$ \\
 8 &  9 &  7.9 &  19   &   286 & ${\sim}24\,000\times$ \\
10 & 11 &  8.9 &  30   &   524 & --- \\
12 & 13 & 15   &  49   &   790 & --- \\
14 & 15 & 24   &  74   &  1075 & --- \\
16 & 17 & 28   & 102   &  1404 & --- \\
18 & 19 & 39   & 144   &  1769 & --- \\
20 & 21 & 47   & 192   &  2179 & --- \\
22 & 23 & 60   & 247   &  2654 & --- \\
24 & 25 & 78   & 344   &  3371 & --- \\
\hline
\end{tabular}
\end{table}

\section{Application example}
\label{sec:application-example}

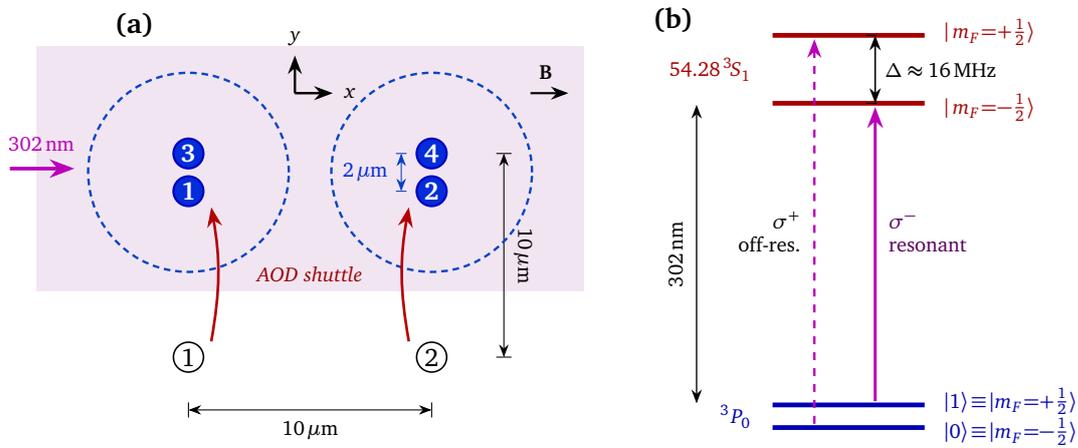
\begin{figure*}[htbp]
\centering
\begin{minipage}[c]{0.44\textwidth}
\centering
\begin{tikzpicture}[
    atom/.style={circle, draw=blue!75!black, line width=0.8pt,
                 fill=blue!65!cyan, text=white, font=\small\bfseries,
                 minimum size=11.26pt, inner sep=0pt},
    atomdim/.style={circle, draw=black, line width=0.6pt,
                    fill=white, text=black, font=\small,
                    minimum size=11.26pt, inner sep=0pt},
    dimline/.style={<->, >=Stealth, black, thin},
    dimtick/.style={black, thin},
    shuttlearrow/.style={->, >=Stealth, red!70!black, line width=1.0pt},
    beamarrow/.style={->, >=Stealth, violet!70!black, line width=1.4pt},
    blockadeline/.style={draw=cyan!55!blue, dash pattern=on 2.5pt off 1.8pt,
                         line width=0.9pt},
    >=Stealth
  ]

  \def\xL{1}        \def\xR{4.2}
  \def\yRtwo{2.7}   
  \def\yRone{0}     
  \def\yRgate{2.2}  

  \fill[violet!25!white, opacity=0.40]
    (-1.0, {2.5-1.62}) rectangle (6.2, {2.5+1.62});

  \node[anchor=north west, font=\bfseries] at (-0.1, 4.7) {(a)};

  \draw[dimline] (\xL,-0.7) -- (\xR,-0.7);
  \draw[dimtick] (\xL,-0.6) -- (\xL,-0.8);
  \draw[dimtick] (\xR,-0.6) -- (\xR,-0.8);
  \node[font=\scriptsize, black, below] at (2.6,-0.72)
    {$10\,\mu\text{m}$};

  \draw[dimline] (\xR+0.95, \yRone) -- (\xR+0.95, \yRtwo);
  \draw[dimtick] (\xR+0.85, \yRone) -- (\xR+1.05, \yRone);
  \draw[dimtick] (\xR+0.85, \yRtwo) -- (\xR+1.05, \yRtwo);
  \node[font=\scriptsize, black, rotate=-90, anchor=south]
    at (\xR+0.95, 1.35) {$10\,\mu\text{m}$};

  \draw[dimline, cyan!50!blue]
    (\xR-0.4, \yRgate) -- (\xR-0.4, \yRtwo);
  \draw[dimtick, cyan!50!blue] (\xR-0.47, \yRgate)--(\xR-0.33, \yRgate);
  \draw[dimtick, cyan!50!blue] (\xR-0.47, \yRtwo)--(\xR-0.33, \yRtwo);
  \node[font=\scriptsize, cyan!35!blue, left]
    at (\xR-0.45, {(\yRgate+\yRtwo)/2}) {$2\,\mu\text{m}$};

  \draw[blockadeline] (\xL, {(\yRgate+\yRtwo)/2}) circle (1.32);
  \draw[blockadeline] (\xR, {(\yRgate+\yRtwo)/2}) circle (1.32);

  \node[atom] (A3) at (\xL, \yRtwo) {3};
  \node[atom] (A4) at (\xR, \yRtwo) {4};

  \node[atomdim] (A1i) at (\xL, \yRone) {1};
  \node[atomdim] (A2i) at (\xR, \yRone) {2};

  \node[atom] (A1g) at (\xL, \yRgate) {1};
  \node[atom] (A2g) at (\xR, \yRgate) {2};

  \draw[shuttlearrow] ([xshift=3mm]A1i.north)
    to[out=80, in=-80] ([xshift=3mm]A1g.south);
  \draw[shuttlearrow] ([xshift=-3mm]A2i.north)
    to[out=100, in=-100] ([xshift=-3mm]A2g.south);
  \node[font=\scriptsize\itshape, red!60!black]
    at (2.6, 1.1) {AOD shuttle};

  \draw[beamarrow, violet!145!black] (-1.35, 2.5) -- (-0.5, 2.5);
  \node[font=\scriptsize, violet!145!black, anchor=south west]
    at (-1.5, 2.6) {$302\,\text{nm}$};

  \draw[beamarrow, black, line width=0.9pt] (2.4, 3.5) -- (2.9, 3.5)
    node[right, font=\scriptsize] {$x$};
  \draw[beamarrow, black, line width=0.9pt] (2.4, 3.5) -- (2.4, 4.0)
    node[above, font=\scriptsize] {$y$};

  \draw[beamarrow, line width=0.8pt, black]
    (5.5, 3.5) -- (6.0, 3.5);
  \node[font=\scriptsize, black, below]
    at (5.7, 4.0) {$\mathbf{B}$};

\end{tikzpicture}
\end{minipage}%
\hfill
\begin{minipage}[c]{0.48\textwidth}
\centering
\begin{tikzpicture}[
    qlevel/.style={line width=1.8pt, draw=blue!70!black},
    rlevel/.style={line width=1.8pt, draw=red!65!black},
    qtext/.style={font=\scriptsize, blue!70!black},
    rtext/.style={font=\scriptsize, red!65!black},
    resonant/.style={->, >=Stealth, line width=1.1pt, violet!145!black},
    restext/.style={font=\scriptsize, violet!70!black},
    offres/.style={->, >=Stealth, dashed, line width=0.9pt, violet!145!black},
    zeemanarrow/.style={<->, >=Stealth, black, line width=0.6pt},
    gaparrow/.style={<->, >=Stealth, black, line width=0.5pt},
    dimtick/.style={black, thin},
  ]

  \def\xLa{0}    \def\xLb{2.0}
  \def\yQlo{0.1}   \def\yQhi{0.4}
  \def\yRlo{4.4} \def\yRhi{5.3}

  \node[anchor=north west, font=\bfseries] at (-1.7, 5.8) {(b)};

  \draw[qlevel] (\xLa, \yQlo) -- (\xLb, \yQlo);
  \draw[qlevel] (\xLa, \yQhi) -- (\xLb, \yQhi);
  \draw[rlevel] (\xLa, \yRlo) -- (\xLb, \yRlo);
  \draw[rlevel] (\xLa, \yRhi) -- (\xLb, \yRhi);

  \node[qtext, left]
    at (\xLa-0.15, {(\yQlo+\yQhi)/2}) {$^3P_0$};
  \node[rtext, left]
    at (\xLa-0.15, {(\yRlo+\yRhi)/2}) {$54.28\,^3\!S_1$};

  \node[qtext, right]
    at (\xLb+0.12, \yQlo-0.05) {$|0\rangle\!\equiv\!|m_F{=}{-}\tfrac{1}{2}\rangle$};
  \node[qtext, right]
    at (\xLb+0.12, \yQhi+0.05) {$|1\rangle\!\equiv\!|m_F{=}{+}\tfrac{1}{2}\rangle$};
  \node[rtext, right]
    at (\xLb+0.12, \yRlo-0.05) {$|\,m_F{=}{-}\tfrac{1}{2}\rangle$};
  \node[rtext, right]
    at (\xLb+0.12, \yRhi+0.05) {$|\,m_F{=}{+}\tfrac{1}{2}\rangle$};

  \draw[resonant] (1.35, \yQhi+0.05) -- (1.35, \yRlo-0.05);
  \node[restext, right, align=left]
    at (1.4, {(\yQlo+\yRhi)/2})
    {$\sigma^-$\\[-1pt]{\scriptsize resonant}};

  \draw[offres] (0.55, \yQlo+0.05) -- (0.55, \yRhi-0.05);
  \node[font=\scriptsize, left, black, align=right]
    at (0.5, {(\yQlo+\yRhi)/2})
    {$\sigma^+$\\[-1pt]{\scriptsize off-res.}};

  \draw[gaparrow] (-1.0, \yQhi+0.02) -- (-1.0, \yRlo-0.02);
  \node[font=\scriptsize, rotate=90, anchor=south]
    at (-1.05, {(\yQhi+\yRlo)/2}) {$302\,\text{nm}$};

  \draw[zeemanarrow] (1.35, \yRlo) -- (1.35, \yRhi);
  \draw[dimtick] (1.23, \yRlo)--(1.37, \yRlo);
  \draw[dimtick] (1.23, \yRhi)--(1.37, \yRhi);
  \node[font=\scriptsize, right, black, align=left]
    at (1.35, {(\yRlo+\yRhi)/2})
    {$\Delta \approx 16\,$MHz};

\end{tikzpicture}
\end{minipage}
    \caption{Setup for the logical Bell state generation protocol with four atoms. \textbf{(a)} Geometry depicting the positions of the atoms and the Rydberg excitation beam. During the protocol, atoms 1 and 2 are shuttled from their original positions (open circles) to within $2\,\mu\text{m}$ of atoms 3 and 4. Dashed circles depict the Rydberg blockade radius ($r_b = 4.9\,\mu\text{m}$) around each qubit pair. The horizonal separation is $10\,\mu\text{m}$ throughout to reduce crosstalk between pairs. After applying entangling gates, atoms 1 and 2 are returned to their initial positions. The violet shaded band indicates the $302\,\text{nm}$ beam cross-section (waist $w_0 = 12\,\mu\text{m}$), which illuminates the gate-position atoms. \textbf{(b)} Level structure of the ytterbium-171 nuclear-spin qubit (blue lines) and the state-selective coupling of $|1\rangle$ to the $|r,m_F{=}{-}1/2\rangle$ Rydberg state via $\sigma^-$ polarized light. An additional $\sigma^+$ polarization component couples $|0\rangle$ to $|r,m_F{=}{+}1/2\rangle$ but is detuned by $\Delta \approx 16.1\,\text{MHz}$.}
    \label{fig:422_system}
\end{figure*}

The capabilities of \atomtwin{} are most clearly demonstrated by a simulation that exercises the full framework: realistic atomic level structure, polarized laser fields, finite-temperature atomic motion, AOD tweezer shuttling, distance-dependent Rydberg interactions, and amplitude-shaped entangling pulses, all derived from physical parameters without building the Hamiltonian by hand. Inspired by recent experimental demonstrations~\cite{reichardt2024,bedalov2024,lib2026}, this section describes an end-to-end simulation of logical Bell state preparation in the $[[4,2,2]]$ quantum error-detecting code, realized on four ytterbium-171 atoms in a $2\times2$ AOD-generated tweezer array (depicted in Fig.~\ref{fig:422_system}); intended as a demonstration of the modeling workflow and expressiveness of the framework, rather than a fully validated prediction of experimental performance. We focus on the state preparation circuit which is sufficient to understand the impact of physical parameters on error probabilities, however we note a complete error-detection protocol would additionally require syndrome measurements to identify and flag errors. We proceed in three steps: characterize isolated CZ gates at a fixed gate separation (Sec.~\ref{sec:422-cz-gate}), introduce the $[[4,2,2]]$ code and its encoding circuit (Sec.~\ref{sec:422-code}), and simulate the full encoding protocol including atom shuttling and dynamical decoupling to evaluate its error-detecting properties (Sec.~\ref{sec:422-model}). The source of this simulation is provided as self-contained Julia scripts in the supplemental code repository. 

\subsection{Qubit encoding and Rydberg gate mechanism}
\label{sec:422-system}

Qubits are encoded in the $^3P_0$ nuclear spin of $^{171}$Yb: $|0\rangle = |m_F = -\tfrac{1}{2}\rangle$ and $|1\rangle = |m_F = +\tfrac{1}{2}\rangle$ (Fig.~\ref{fig:422_system}b) and trapped at the clock-magic wavelength $759\,\text{nm}$ by a two-dimensional AOD tweezer array~\cite{Ma2023,peper2024}. Because $^3P_0$ has $J = 0$, the electronic Zeeman contribution vanishes; the qubit splitting is set entirely by the nuclear Zeeman shift, ${\sim}9.5\,\text{kHz/G}$.

Entanglement between the qubit states is generated by a linearly polarized $302\,\text{nm}$ UV laser coupling $^3P_0$ to the $54s\,^3S_1$ Rydberg manifold (effective principal quantum number $n^* = 54.28$~\cite{peper2024}). A static magnetic field $\mathbf{B} \parallel \hat{x}$ splits the two $m_F$ sublevels of the Rydberg state by $\Delta = g_\text{eff}\,\mu_B\,B/h = 16.1\,\text{MHz}$ at $B = 4.88\,\text{G}$, where the effective $g$-factor $g_\text{eff} = 2.357$ accounts for configuration mixing with nearby $J=1$ states~\cite{peper2024}. The UV beam propagates along $\hat{x}$ so that in the quantization frame set by $\mathbf{B}$, its polarization decomposes into equal $\sigma^+$ and $\sigma^-$ components. The laser frequency is tuned to bring $|r,\,m_F = -1/2\rangle$ on resonance, coupling $|1\rangle$ resonantly to the Rydberg state with $\Omega_R/2\pi = 2.5\,\text{MHz}$ at $P=20\,$mW and beam radius $w=12\,\mu$m. The $\sigma^+$ component couples $|0\rangle$ to $|r,\,m_F = +1/2\rangle$, but this transition is detuned by $\Delta \approx 16.1\,\text{MHz}$, leaving $|0\rangle$ mostly dark and realizing the qubit-selective blockade required for a CZ gate. The van der Waals interaction at the $2\,\mu\text{m}$ gate separation is $V/2\pi = C_6 / (2\,\mu\text{m})^6 \approx 531\,\text{MHz}$, giving $V/\Omega_R \approx 212$: well within the strong blockade regime. 

\subsection{Time-optimal CZ gate}
\label{sec:422-cz-gate}

\begin{figure}[htbp]
    \centering
    \includegraphics[width=\linewidth]{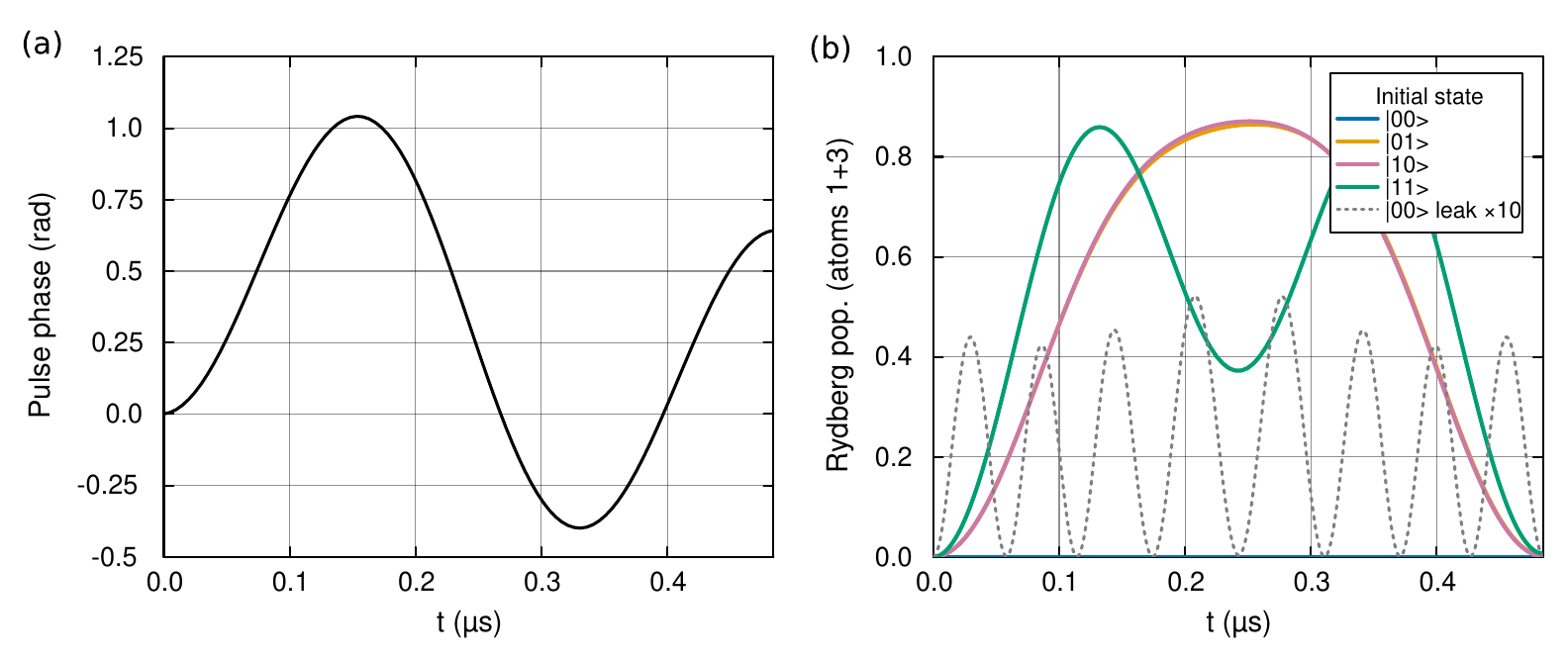}
    \caption{Time-optimal CZ gate simulation.
    \textbf{(a)} Phase envelope $\phi(t)$ of the fixed-amplitude Rydberg drive over $T_\text{gate} \approx 0.48\,\mu\text{s}$, from the data provided with Ref.~\cite{jandura2022}.
    \textbf{(b)} Rydberg state population $P_r(t)$ averaged over both blockaded atom pairs, for all four computational basis inputs. In the $|11\rangle$ case, blockade prevents the double excitation resulting in collective dynamics that differ from the $|01\rangle,|10\rangle$ cases. The $\sigma^+$ leakage into the off-resonant $|r,\,m_F{=}{+}\tfrac{1}{2}\rangle$ sublevel (dashed, $\times 10$) undergoes rapid oscillations at the Zeeman splitting frequency ($\Delta/2\pi \approx 16.1\,\text{MHz}$) but returns nearly to zero at $t = T_\text{gate}$, leaving negligible residual leakage despite the linearly polarized drive.}
    \label{fig:cz_gate}
\end{figure}

The two-qubit entangling gate follows the time-optimal protocol of Jandura and Pupillo~\cite{jandura2022}, which achieves a CZ operation using a fixed-amplitude Rydberg drive with a time-dependent phase $\phi(t)$ (Fig.~\ref{fig:cz_gate}a). The pulse couples $|1\rangle$ resonantly to $|r,\,m_F=-\tfrac{1}{2}\rangle$ via the $\sigma^-$ beam component with $\Omega_-/2\pi = 2.5\,\text{MHz}$, giving $T_\text{gate} = 7.612/\Omega_- \approx 0.48\,\mu\text{s}$. The phase envelope is reconstructed from numerically tabulated values~\cite{jandura2022} using piecewise-constant interpolation. By design, each basis state returns to zero Rydberg population at $t = T_\text{gate}$ (Fig.~\ref{fig:cz_gate}b): $|00\rangle$ is decoupled from the drive; $|01\rangle$ and $|10\rangle$ undergo single-atom oscillations; and $|11\rangle$, confined by blockade to the symmetric Dicke subspace $\{|11\rangle,\,(|1r\rangle+|r1\rangle)/\sqrt{2}\}$, accumulates a distinct collective phase. The pulse is optimized so that the conditional phase $\varphi_{11} - \varphi_{10} - \varphi_{01} + \varphi_{00} = \pi$, realizing the entangling gate. Our simulation reproduces the Rydberg state populations of Ref.~\cite{jandura2022}; deviations from the ideal gate are isolated and quantified in the error budget below.

Unlike the original gate analysis~\cite{jandura2022}, the simulation naturally includes seven experimentally relevant error sources: Doppler dephasing, beam inhomogeneity, spontaneous decay, finite blockade, inter-pair crosstalk, Zeeman dephasing, and $\sigma^+$ coupling. Building it proceeds in three steps.

\textit{Step 1: physical model.} \cbox{HyperfineManifold} specifies quantum numbers and $g$-factors from which \atomtwin{} derives the Hilbert space; \cbox{Ytterbium171Atom} samples thermal velocities from the Maxwell--Boltzmann distribution; \cbox{GeneralGaussianBeam} carries the full beam geometry and \cbox{rabi\_frequencies} decomposes its polarization into $\sigma^+/\sigma^-/\pi$ components in the frame set by $\mathbf{B}$.

\begin{lstlisting}[language=julia, label={lst:model}]
met     = HyperfineManifold(1//2, 0; label="3P0",     g_F=-0.00067875)
rydberg = HyperfineManifold(1//2, 0; label="54.28S1", g_F=g_eff)
r, leak = rydberg[-1//2], rydberg[+1//2]
atoms   = [Ytterbium171Atom(; levels=[met..., rydberg...],
                               v_init=maxwellboltzmann(T=3µK)) for _ in 1:4]
beam    = GeneralGaussianBeam(302nm, w_ryd, w_ryd, P_ryd, k_ryd, pol_ryd)
Ω_π, Ω_p, Ω_m = rabi_frequencies(beam; q_axis=B_vec, d_red=d_eff)
\end{lstlisting}

\noindent\textit{Step 2: system assembly.} \cbox{add\_zeeman\_detunings!} derives level shifts from the $g$-factors and field magnitude; \cbox{add\_coupling!} registers a coherent drive; passing a \cbox{Beam} makes the coupling position-dependent, with the complex Rabi amplitude recomputed each timestep from the atom's instantaneous position relative to the beam; \cbox{add\_vdwinteraction!} registers a $C_6/r^6$ interaction operator recomputed from the instantaneous atomic separation; \cbox{add\_decay!} installs quantum jump operators for spontaneous emission from the Rydberg state.

\begin{lstlisting}[language=julia, label={lst:system}]
sys = System(atoms, [tweezer])
for atom in atoms
    add_zeeman_detunings!(sys, atom, met,     B=B)
    add_zeeman_detunings!(sys, atom, rydberg, B=B, delta=Δ_ryd)
end
add_vdwinteraction!(sys, (atoms[1],atoms[3]), (r,r)=>(r,r), C6)
add_vdwinteraction!(sys, (atoms[2],atoms[4]), (r,r)=>(r,r), C6)
r = vcat([add_coupling!(sys, atom, met=>rydberg; beam=beam,
                Ω_π=Ω_π, Ω_p=Ω_p, Ω_m=Ω_m, active=false) for atom in atoms]...)
for atom in atoms; add_decay!(sys, atom, rydberg=>met, Γ_r); end
\end{lstlisting}

\noindent\textit{Step 3: pulse and error budget.} The gate is a single amplitude-modulated pulse applied in parallel to all atoms in the beam. For the error budget, \cbox{build\_system} is a helper that constructs the system from the same physical parameters but with each experimental imperfection toggled via boolean keyword flags (e.g.\ \cbox{doppler\_dephasing=true}, \cbox{spontaneous\_decay=false}); rebuilding with one flag set at a time isolates each source. \cbox{play} is then called over all four computational inputs (\ilcode{initial\_state=[ket1,q1,ket3,q1]}); atoms 2 and 4 are initialized in $|1\rangle$ throughout, so they are always within range of the Rydberg drive and contribute crosstalk effects on the gate qubits (atoms 1 and 3). The five metrics of Table~\ref{tab:cz_budget} are extracted from the returned wavefunctions. Stochastic sources are averaged over $n = 10\,000$ WFMC trajectories.

\begin{lstlisting}[language=julia, label={lst:cz_gate}]
for (src, stochastic) in error_sources       # one source at a time
    flags = merge(all_off, NamedTuple{(src,)}((true,)))
    (; sys, r) = build_system(; flags...)
    seq = Sequence(dt; downsample=10)
    @sequence seq begin
        Pulse(r, T_gate; amplitudes=ryd_amplitudes, interp=:piecewise_constant)
    end
    for (ket1,ket3) in [(q0,q0),(q0,q1),(q1,q0),(q1,q1)]
        out = play(sys, seq; initial_state=[ket1,q1,ket3,q1], shots=N)
    end
end
\end{lstlisting}

\begin{table*}[htbp]
    \centering
    \caption{CZ gate error budget computed with \atomtwin{}. Each error source is activated independently; the final row enables all simultaneously. $\langle P_\text{comp}\rangle$: mean probability to return to the correct computational basis state, averaged over all four inputs. $\delta\varphi_\text{sq}$ and $\delta\varphi_\text{CZ}$: mean phase deviations from the ideal values $\theta_\text{sq} = 2.1663\,\text{rad}$ and $\pi$; the single-qubit bias is correctable by $R_Z$ rotations. $\epsilon_\text{sq} = 1-|\langle e^{i\delta\varphi_\text{sq}}\rangle|$ and $\epsilon_\text{CZ} = 1-|\langle e^{i\delta\varphi_\text{CZ}}\rangle|$: stochastic phase coherence errors; reported only for the stochastic sources ($n=10\,000$ trajectories each). $F_e$: entanglement fidelity estimated from Eq.~\eqref{eq:agf}; off-diagonal elements of the error process are negligible, confirming the diagonal approximation. Zeeman dephasing assumes a static, uniform magnetic field and is therefore deterministic. Spontaneous decay uses the measured Rydberg lifetime $\tau_r = 56\,\mu\text{s}$~\cite{peper2024}. The total row enables all sources simultaneously and includes cross-term contributions (e.g., atomic motion amplifies the effects of beam inhomogeneity and finite blockade).}
    \label{tab:cz_budget}
    \begin{tabular}{lcccccc}
        \hline\hline
        Error source & $\langle P_\text{comp}\rangle$ & $\delta\varphi_\text{sq}$ (rad) & $\epsilon_\text{sq}$ & $\delta\varphi_\text{CZ}$ (rad) & $\epsilon_\text{CZ}$ & $F_e$ \\
        \hline
        Ideal (all off)      & 1.0000 & $+0.0001$ & ---      & $+0.0002$ & ---      & 1.0000 \\
        \hline
        Doppler dephasing    & 0.9981 & $+0.0008$ & $0.0039$ & $+0.0007$ & $0.0090$ & 0.9897 \\
        Spontaneous decay    & 0.9992 & $+0.0012$ & $0.0028$ & $-0.0007$ & $0.0068$ & 0.9931 \\
        Beam inhomogeneity   & 0.9995 & $-0.0124$ & ---      & $+0.0053$ & ---      & 0.9995 \\
        Finite blockade      & 1.0000 & $+0.0001$ & ---      & $+0.0093$ & ---      & 1.0000 \\
        Inter-pair crosstalk & 0.9999 & $-0.0314$ & ---      & $+0.0297$ & ---      & 0.9998 \\
        Zeeman dephasing     & 1.0000 & $+0.0106$ & ---      & $+0.0038$ & ---      & 1.0000 \\
        $\sigma^+$ coupling  & 1.0000 & $-0.2916$ & ---      & $+0.0002$ & ---      & 1.0000 \\
        \hline
        Total (all on)       & 0.9964 & $-0.3216$ & $0.0069$ & $+0.0487$ & $0.0181$ & 0.9801 \\
        \hline\hline
    \end{tabular}
\end{table*}

The results in Table~\ref{tab:cz_budget} show that the ideal baseline (all error sources off) reproduces the target gate almost perfectly, with a small two-qubit CZ phase bias $|\delta\varphi_{2}| = 0.2\,\mathrm{mrad}$ due to the finite interaction strength used in the simulation. Doppler dephasing produces the largest loss from the initial state averaged over all four computational states ($\langle P_{\mathrm{comp}}\rangle = 0.9981$), followed by spontaneous decay ($0.9992$); all other sources leave populations essentially unchanged. The most significant coherent phase error is the $\sigma^+$ coupling, which imprints a single-qubit phase bias of $-0.29\,\mathrm{rad}$ (correctable with $R_z$ rotations). Interestingly, it has negligible effect on the populations due to near-complete refocusing of off-resonant Rabi oscillations for $|B| = 4.88\,$G. Crosstalk from the second pair produces a CZ phase bias of $0.03\,\mathrm{rad}$ that depends on the four-qubit state and cannot be removed by single-qubit rotations alone. 

Stochastic gate errors are dominated by Doppler dephasing ($\epsilon_\text{CZ} = 0.0090$, $\epsilon_\text{sq} = 0.0039$) and spontaneous decay ($\epsilon_\text{CZ} = 0.0068$, $\epsilon_\text{sq} = 0.0028$). The larger CZ dephasing relative to population error arises because phase coherence ($\epsilon_\text{CZ}$) is degraded by any decay event, including those that return the atom to a qubit state with a random phase, whereas population error ($1 - \langle P_\text{comp}\rangle$) counts only events that remove amplitude from the target state averaged over all four inputs. From these metrics we estimate the entanglement fidelity~\cite{nielsen2002} in the diagonal approximation (off-diagonal elements of the error process matrix are confirmed to be negligible in our simulation):
\begin{equation}
    F_e \approx \frac{1}{d^2}\left|\sum_j \langle\sqrt{P_j}\,e^{i\delta\varphi_j}\rangle\right|^2 = 0.980,
    \label{eq:agf}
\end{equation}
where $d=4$ is the Hilbert space dimension of the two-qubit gate, $j$ labels the four computational basis inputs, and $\delta\varphi_j$ is the per-trajectory phase deviation from ideal after removing calibratable single-qubit biases.

\subsection{The \texorpdfstring{$[[4,2,2]]$}{[[4,2,2]]} code and encoding circuit}
\label{sec:422-code}

\begin{figure*}[htbp]
\centering
\begin{minipage}[b]{0.32\linewidth}
\centering
{\small (a) Encoding circuit}\\[6pt]
\begin{quantikz}[column sep=5mm, row sep=7mm]
  \lstick{$q_1$} & \gate{H} & \ctrl{2} & \qw      & \qw \\
  \lstick{$q_2$} & \gate{H} & \qw      & \ctrl{2} & \qw \\
  \lstick{$q_3$} & \qw      & \targ{}  & \qw      & \qw \\
  \lstick{$q_4$} & \qw      & \qw      & \targ{}  & \qw
\end{quantikz}
\end{minipage}%
\hspace{6mm}%
\begin{minipage}[b]{0.62\linewidth}
\centering
{\small (b) Hardware-compiled circuit}\\[1pt]
\begin{quantikz}[column sep=3.5mm, row sep=7mm]
	\lstick{$q_1$} & \gate{H}
	& [-3mm] \qw\slice{} & [10mm] \ctrl{2}
	& \qw & \qw & \gate{X}
	& [-3mm] \qw\slice{} & [10mm] \gate{X} & \qw & \qw\\
	\lstick{$q_2$} & \gate{H}
	& \qw & \qw
	& \ctrl{2} & \qw & \gate{X}
	& \qw & \gate{X} & \qw & \qw\\
	\lstick{$q_3$} & \gate{H}
	& \qw & \control{}
	& \qw & \qw & \gate{X}
	& \qw & \gate{X} & \gate{H} & \qw\\
	\lstick{$q_4$} & \gate{H}
	& \qw & \qw
	& \control{} & \qw & \gate{X}
	& \qw & \gate{X} & \gate{H} & \qw
  \end{quantikz}
\end{minipage}
\caption{Encoding circuits for the $[[4,2,2]]$ logical Bell state.
\textbf{(a)} Logical circuit: Hadamard gates on $q_1$ and $q_2$ prepare
$|{+}{+}{0}{0}\rangle$; two simultaneous CNOT gates on pairs $(1,3)$ and $(2,4)$ produce the state $|\Phi^+_L\rangle = (|00_L\rangle + |11_L\rangle)/\sqrt{2}$ defined in the text. \textbf{(b)} Hardware-compiled circuit using native operations for neutral atom QPUs.
$H = R_z(\pi/2)\cdot R_x(\pi/2)\cdot R_z(\pi/2)$ is the
hardware decomposition of the Hadamard gate; and $X = R_x(\pi)$.
The $CZ$ gates act simultaneously on row pairs $(1,3)$ and $(2,4)$,
which are brought within the Rydberg blockade radius by moving the atoms (vertical dashed lines).
Two $X$ echo pulses sandwich the return move, canceling the dynamical phase accumulated during transport.}
\label{fig:422_circuit}
\end{figure*}
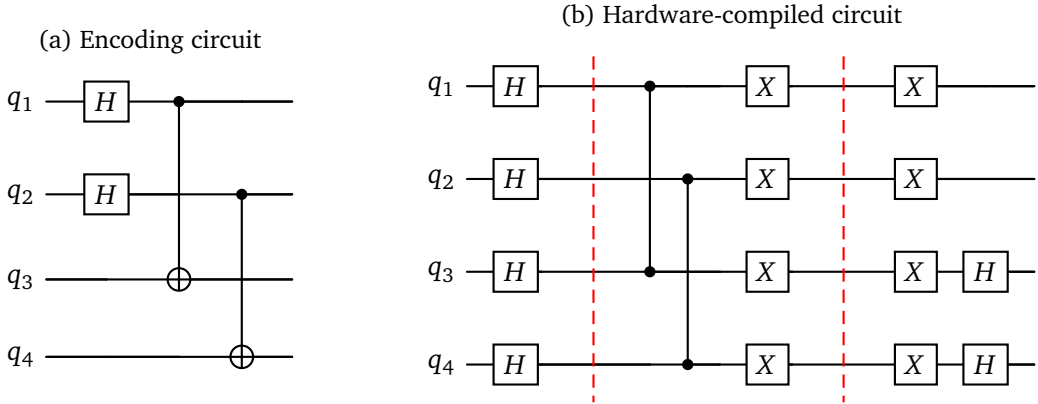

The $[[4,2,2]]$ code encodes two logical qubits in four physical qubits, with stabilizer generators $S_X = XXXX$ and $S_Z = ZZZZ$. This code is the smallest non-trivial quantum error-detecting code that detects (but cannot correct) any single-qubit Pauli error~\cite{gottesman1997}. The four logical codewords are:
\begin{align}
    |00_L\rangle &= \tfrac{1}{\sqrt{2}}\bigl(|0000\rangle + |1111\rangle\bigr), &
    |01_L\rangle &= \tfrac{1}{\sqrt{2}}\bigl(|0011\rangle + |1100\rangle\bigr), \nonumber\\
    |10_L\rangle &= \tfrac{1}{\sqrt{2}}\bigl(|0110\rangle + |1001\rangle\bigr), &
    |11_L\rangle &= \tfrac{1}{\sqrt{2}}\bigl(|0101\rangle + |1010\rangle\bigr).
    \label{eq:422_codewords}
\end{align}
The target state of our simulation is the logical Bell state $|\Phi^+_L\rangle = (|00_L\rangle + |11_L\rangle)/\sqrt{2}$, which in the physical basis reads
\begin{equation}
    |\Phi^+_L\rangle = \tfrac{1}{2}\bigl(|0000\rangle + |0101\rangle + |1010\rangle + |1111\rangle\bigr).
    \label{eq:logical_bell}
\end{equation}
This state has equal weight on four computational basis states and is prepared by the logical circuit $H(1,2) \to \text{CX}(1,3)\cdot\text{CX}(2,4)$ (Fig.~\ref{fig:422_circuit}a), with $CX$ acting on the blockade-coupled pairs (1,3) and (2,4). In terms of native neutral-atom gates this becomes $H^{\otimes 4} \to \text{CZ}(1,3)\cdot\text{CZ}(2,4) \to H(3,4)$ (Fig.~\ref{fig:422_circuit}b), via $\text{CX}(c,t) = H(t)\cdot\text{CZ}(c,t)\cdot H(t)$ with target Hadamards absorbed into adjacent layers. Each $H$ can be decomposed into a composite $R_x\cdot R_z$ rotation and $R_z$ pulses provide additional phase corrections for the $CZ$ gates. Gates within each layer are applied in parallel: a single global Rydberg pulse implements both $CZ$ gates; both the $H^{\otimes 4}$ and $X$ layers can be performed with global pulses ($X$ implemented as $R_x(\pi)$), whereas the final $H(3,4)$ layer requires row-resolved spatial addressing.

Single-qubit gates ($R_x$, $R_z$) can in principle be modeled at the level of the full Raman coupling and hyperfine structure, as discussed in the examples packaged with \atomtwin{}. Here we adopt a simplified model: $R_x$ as an effective resonant coupling between $|0\rangle$ and $|1\rangle$ and $R_z$ as a differential energy offset on $|1\rangle$, both at $1\,\text{MHz}$, since single-qubit errors are not the dominant source of infidelity for this circuit and adding auxiliary hyperfine states would substantially expand the Hilbert space.

\subsection{Logical Bell state generation}
\label{sec:422-model}

The full protocol uses the same four-atom system as the gate characterization but starts from the experimental geometry ($y_\text{sep} = 10\,\mu\text{m}$), so atoms must be shuttled $8\,\mu\text{m}$ before the gate and returned afterwards. The $CZ$ analysis fixed the atoms at the gate separation and omitted single-qubit operations; the encoding sequence adds \cbox{MoveRow} shuttling, dynamical decoupling echoes, and row-addressed Hadamards. These introduce additional error channels: Doppler dephasing during transport, shot-to-shot variation in the Rabi frequency and van der Waals strength, and residual position uncertainty at the gate time. Two couplings not present in Sec.~\ref{sec:422-cz-gate} are added: a resonant $|0\rangle\leftrightarrow|1\rangle$ coupling (\cbox{sq}) and a differential energy offset on $|1\rangle$ (\cbox{det}) for the simplified $R_x$ and $R_z$ gates. Gate operations are then composed as functions returning \cbox{Pulse} lists. \cbox{RZ} is a timed detuning on $|1\rangle$; \cbox{CZ} appends $R_Z$ phase corrections after the Rydberg drive.

\begin{lstlisting}[language=julia, label={lst:helpers}]
RZ(t, ph) = [Pulse(det[t], mod(ph, 2π)/Δ_det)]
H(t)      = [RZ(t, π/2)..., Pulse(sq[t], T_π/2), RZ(t, π/2)...]
X(t)      = [Pulse(sq[t], T_π)]
CZ()      = [Pulse(r, T_gate; amplitudes=amps, interp=:piecewise_constant), RZ(1:4, θ_cz)...]
\end{lstlisting}

\noindent The \cbox{@sequence} macro assembles these into a hardware timeline. \cbox{MoveRow} ramps an AOD drive frequency over duration $T_\text{move}$, translating row~1 by $8\,\mu\text{m}$ while integrating quantum and motional dynamics simultaneously. The total sequence duration is $2T_\text{move} + T_\text{gate} + 4T_\pi \approx 164\,\mu\text{s}$, dominated by transport; the surrounding \cbox{X} echoes cancel static dephasing without disturbing the CZ-acquired phase.

\begin{lstlisting}[language=julia, label={lst:seq}]
(; sys, r, atoms, tweezer) = build_system(y_start=ysep)
seq = Sequence(dt; downsample=1000)
@sequence seq begin
    Wait(0.1µs)
    H(1:4)                                                      # global Hadamard
    MoveRow(tweezer, 1,  (ysep-dy_gate)/µm*MHz, T_move; dt=10dt)  # shuttle
    CZ()                                                        # entangling gate
    X(1:4)                                                      # decoupling echo
    MoveRow(tweezer, 1, -(ysep-dy_gate)/µm*MHz, T_move; dt=10dt)  # return
    X(1:4)                                                      # decoupling echo
    H(3:4)                                                      # row Hadamard
end
\end{lstlisting}

The simulation runs for $n_\text{shots} = 100$ independent Monte Carlo trajectories using the full $4^4 = 256$-dimensional Hilbert space. The dominant basis states in the final wavefunction are the four target components $|0000\rangle$, $|0101\rangle$, $|1010\rangle$, $|1111\rangle$, each with probability ${\approx} 0.24$--$0.25$, with the residual ${\approx}6\%$ population distributed across states with odd parity (outside the logical subspace). Full results are summarized in Table~\ref{tab:422_fidelity}.

\begin{table}[htbp]
    \centering
    \caption{Fidelity metrics for the $[[4,2,2]]$ logical Bell state simulation (mean over 100 shots, $T = 3\,\mu\text{K}$). $F_\text{raw}$ is the overlap with the ideal target state~\eqref{eq:logical_bell}; $P_\text{even}$ is the probability of an even-parity measurement outcome (Z-stabilizer post-selection); $F_\text{post}$ is the fidelity conditioned on $P_\text{even}$; $F_\text{syndrome}$ is the fidelity conditioned on the full syndrome (Z and X stabilizers).}
    \label{tab:422_fidelity}
    \begin{tabular}{lc}
        \hline
        Metric & Value \\
        \hline
        Raw fidelity $F_\text{raw}$ & $0.942$ \\
        Z-stabilizer acceptance $P_\text{even}$ & $0.968$ \\
        Post-selected fidelity $F_\text{post}$ & $0.973$ \\
        Full-syndrome fidelity $F_\text{syndrome}$ & $0.9996$ \\
        \hline
    \end{tabular}
\end{table}

The raw fidelity $F_\text{raw} = 0.942$ reflects errors that project the state out of the code space, predominantly caused by left-over Rydberg population after the gate and subsequent decay during the final move. Post-selecting on the $Z$ stabilizer ($P_\text{even} = 0.968$) removes many of these events, raising the fidelity to $F_\text{post} = 0.973$; this post-selection is experimentally realizable by measuring the parity of the physical qubit populations. Post-selecting additionally on the $X$ stabilizer (full syndrome) recovers $F_\text{syndrome} = 0.9996$. The near-unity syndrome-corrected fidelity indicates that the dominant error sources are detectable rather than logical errors, without any assumptions on the Clifford or Pauli character of the noise.

At four physical qubits, with a total Hilbert space dimension of $4^4 = 256$, the system is tractable for direct physics-level simulation. For larger circuits, \atomtwin{} can be used to extract gate-level Kraus maps from individual physics simulations. If desired, these can then be composed in circuit-based emulators to scale to much larger protocols.
\section{Conclusion}
\label{sec:conclusion}
We have presented \atomtwin{}, a Julia framework for physics-level digital twins of neutral-atom quantum processors. It sits between circuit-level simulators, which operate on abstract gate sets and effective noise models, and full \textit{ab initio} calculations, which are intractable at the system sizes relevant to near-term devices. By composing physical building blocks---laser beams, magnetic field parameters, hyperfine manifolds, tweezer arrays, and AOD transport---into a simulation that drives the same equations of motion as the hardware, \atomtwin{} lets users connect instruction-level metrics directly to physical parameters without manually assembling the Hamiltonian. Beyond its immediate use as a simulation tool, this approach suggests a path toward shared, physics-native representations of neutral-atom systems, enabling more consistent modeling and direct comparison across the community.

The framework is organized into two cooperating layers. The user-facing \atomtwin{} layer provides component constructors and orchestration logic; the \module{Dynamiq} engine implements highly optimized time integrators. A user specifies a device in physical terms---atomic species, beam geometry, trap configuration, pulse sequence---and \atomtwin{} compiles this description into a directed acyclic graph (DAG) of typed physics nodes, resolves operator dependencies automatically, and produces a pre-allocated \cbox{SimulationJob}. Parameters and noise realizations are represented as typed objects resolved at compile time, so that stochastic sweeps and parametric studies incur no overhead inside the solver loop. The Schrödinger, Lindblad master equation, and Monte Carlo wave function solvers share this compilation pipeline and are selected automatically based on the system definition. 

Performance was tested against \module{QuantumOptics.jl} and \module{QuTiP} on two benchmarks: single-qubit Rabi oscillations with dephasing, for which an exact analytical solution provides an accuracy reference, and collective Rydberg blockade dynamics for $N = 2$--$8$ atoms, which tests scaling under a stiff many-body Hamiltonian. \atomtwin{} reproduces both references to within numerical integration tolerances while running up to $3\times$ faster than \module{QuantumOptics.jl} for the Schrödinger equation, $2$--$3\times$ faster for the master equation, and ${\sim}5\times$ faster (serial) for MCWF simulations. We emphasize, however, that these comparisons depend on solver configuration and the specific problems of interest and should be interpreted as indicative rather than definitive.

The application example demonstrates end-to-end simulation of logical Bell state preparation in the $[[4,2,2]]$ quantum error-detecting code on four ytterbium-171 atoms. Realistic beam polarization, Zeeman sublevels, finite-temperature atomic motion, AOD shuttling, and time-optimal Rydberg gate pulses are combined in a single simulation with a 256-dimensional Hilbert space. The fidelity $F_\text{raw} = 0.942$ rises to $F_\text{syndrome} = 0.9996$ after full syndrome post-selection using only experimentally realistic physical parameters. The CZ gate error budget further illustrates how individual error sources ($\sigma^+$ coupling, inter-pair crosstalk, spontaneous decay, Doppler dephasing) can be isolated and quantified by activating each independently within the same model. 

Several limitations of the current framework should be noted. The numerical integration uses fixed timesteps without adaptive error control, and scalability is constrained by the exponential growth of the Hilbert space. The treatment of atomic motion is semiclassical, and experimental validation against hardware is not included in this work. Quantum motional effects, such as photon recoil, sideband structure and entanglement between motional and internal states are further directions not addressed in the present version. These aspects represent important directions for future development.

The modular DAG architecture provides a clear path for all of these extensions: each new physical process is a new node type that participates in the same compilation and value-resolution protocol as existing nodes. Possible additions include GPU-accelerated trajectory parallelism, additional atomic species and interaction types, and integration with experimental control stacks for closed-loop calibration. \atomtwin{} the benchmark scripts and the application example are included in the repository as starting points for new simulations.

Beyond its current capabilities, it is useful to situate \atomtwin{} within the broader question of what a quantum digital twin for a quantum processor should ultimately be, although there is not yet a consensus definition. A central challenge is that full-scale simulation of a many-qubit processor is computationally intractable at near-term device sizes, even with advanced methods such as tensor networks. As a result, a predictive digital twin cannot rely on a single modeling layer, but instead requires a multiscale architecture: a physics-accurate simulator at the level of hardware primitives such as \atomtwin{}, combined with higher-level models operating at the circuit and application layers, where full physics simulation is no longer feasible.

Borrowing the concept of maturity levels from other engineering disciplines, one can sketch a natural progression of quantum digital twin capabilities. At the base (Level 0) are structural twins: simulations built on simplified or effective Hamiltonians with little direct connection to experimental data, representing the current standard in theory–experiment workflows. \atomtwin{} corresponds to the next stage (Level 1): parameters are informed by experimental knowledge, dominant error channels are modeled from physical parameters in a consistent framework, and the simulator can guide protocol design and the interpretation of experiments within its calibrated domain. Higher levels should introduce increasing degrees of validation and integration with the physical device. A predictive twin (Level 2) should be systematically benchmarked against experimental measurements to achieve quantitative accuracy beyond calibrated observables. An in-the-loop twin (Level 3) incorporates live experimental feedback for continuous parameter updates and adaptive calibration. At the highest levels (Levels 4–5), the twin becomes part of the control stack, enabling closed-loop autonomous calibration, optimization, and, ultimately, adaptation of both model and control strategies. In this sense, \atomtwin{} provides the physics layer required to bridge structural and predictive digital twins, and offers a foundation for the development of closed-loop, hardware-integrated quantum digital twins in the future.


\section*{Acknowledgements}

SW thanks Manuel Morgado and Guido Masella for early discussions that inspired the architecture of \atomtwin{}; Swayangdipta Bera and Amar Bellahsene for testing the package as well as the aQCess and QPerfect teams for input on example usecases. The author used the Claude Opus 4.6 large language model (Anthropic) to assist in drafting portions of the manuscript and refining parts of the code. All technical content, code, and conclusions were independently developed, reviewed, and verified by the author. \atomtwin{} was initially developed as an educational module within the DigiQ project, to give students and researchers hands-on access to a realistic neutral-atom processor model.

\paragraph{Funding information}
This work received support from the European Union's Digital Europe Programme through the DigiQ project (Digitally Enhanced Quantum Technology Master, grant No.\ 101084035), from the European Union's Horizon Europe programme through the EuRyQa project (European infrastructure for Rydberg Quantum Computing, grant No.\ 101070144), and from the French \textit{Programme d'Investissements d'Avenir} through the aQCess project (Atomic Quantum Computing as a Service, Université de Strasbourg / CNRS). The author acknowledges support from the Institut Universitaire de France (IUF).

\paragraph{Code availability}
\atomtwin{} is distributed through the Julia General registry and is open source under the Apache License 2.0. The source code is available at \url{https://github.com/aQCess/AtomTwin.jl}. This paper corresponds to version~v0.1.1.


\begin{appendix}
	\numberwithin{equation}{section}
	\newpage
\section{Getting started}
\label{sec:getting_started}

\subsection{Installation}

\atomtwin{} is a Julia package distributed through the Julia General registry. Julia~1.11 or later is required. The recommended way to install and manage Julia on all platforms is \cbox{juliaup}, available from the official Julia download page (\url{https://julialang.org/downloads}).

To make use of multithreaded trajectory parallelism (Section~\ref{sec:benchmark}), start Julia with the \ilcode{--threads} flag:

\begin{lstlisting}[language=bash]
julia --threads auto
\end{lstlisting}

\noindent To install \atomtwin{}, enter the Julia package manager with \ilcode{]} and add the package:

\begin{lstlisting}[language=julia]
julia> ]
pkg> add AtomTwin
\end{lstlisting}

\noindent The package is then available in any Julia session with \ilcode{using AtomTwin}. For a reproducible project environment, create a dedicated directory and activate a local environment before adding the package:

\begin{lstlisting}[language=bash]
mkdir my_project && cd my_project
julia --project=. --threads auto
\end{lstlisting}

\noindent This creates a \ilcode{Project.toml} that pins the \atomtwin{} version for the project.

\subsection{First simulation}

A minimal \atomtwin{} simulation follows three steps: define the system, build a sequence, and run with \cbox{play}. The example below simulates Rabi oscillations of a two-level atom driven at $\Omega/2\pi = 1\,\text{MHz}$ for $5\,\mu\text{s}$ and records the excited-state population.

\begin{lstlisting}[language=julia]
using AtomTwin

# Step 1: define the system
g, e  = Level(; label = "g"), Level(; label = "e")
atom  = Atom(; levels = [g, e])
system = System(atom)

add_coupling!(system, atom, g => e, 2pi * 1e6)   # Omega/2pi = 1 MHz
add_detector!(system, PopulationDetectorSpec(atom, e; name = "P_e"))

# Step 2: build a sequence
seq = Sequence(1e-9)   # fixed time step: 1 ns
@sequence seq begin
    Wait(5e-6)         # evolve for 5 us
end

# Step 3: run and inspect
out = play(system, seq; initial_state = g)

# out.detectors["P_e"] is a Vector{Float64} of P_e(t)
# out.times is the corresponding time axis
\end{lstlisting}

\noindent Physical unit literals (\cbox{MHz}, \cbox{\textmu m}, \cbox{mW}, etc.) are available via \ilcode{using AtomTwin.Dynamiq.Units} and are used throughout the examples in this paper.

\subsection{Documentation and examples}

Comprehensive documentation, including an API reference and a collection of worked examples, is available at:
\begin{center}
    \url{https://aqcess.github.io/AtomTwin.jl}
\end{center}

\noindent The documentation examples cover the following workflows:

\begin{itemize}
    \item Rabi oscillations with spontaneous decay (Lindblad master equation)
    \item Rabi oscillations with laser phase noise (\cbox{LaserPhaseNoiseModel})
    \item Rabi oscillations with atomic motion (semiclassical, Doppler shift)
    \item Rabi oscillations with static intensity disorder (\cbox{Parameter} sweep)
    \item Rydberg blockade: collective oscillations and superatom regime
    \item Time-optimal Rydberg CZ gate (Jandura--Pupillo pulse)~\cite{jandura2022}
    \item Single-qubit gate tomography (\cbox{process\_tomography}, Choi matrix, PTM)
    \item Electromagnetically induced transparency with dissipation
    \item Ytterbium-171 Raman qubit gate with hyperfine structure
    \item Potassium-39 state preparation with Zeeman sublevels
    \item Atom sorting in a tweezer array using AOD ramps
\end{itemize}

\noindent Each example is provided as a self-contained Julia script. The simulation scripts used to produce the results and figures in this paper are included in the repository under \ilcode{paper/}.

\end{appendix}


\bibliography{bibliography.bib}

\end{document}